\documentclass[12pt,preprint]{aastex}

\newcommand{\Draft}{false}

\usepackage{amsmath}

\usepackage{xspace}

\newcommand{\powersep}{\times}

\newcommand{\g}{{\ensuremath{\mathrm{g}}}\xspace}
\newcommand{\K}{{\ensuremath{\mathrm{K}}}\xspace}
\newcommand{\cm}{{\ensuremath{\mathrm{cm}}}\xspace}

\newcommand{\km}{{\ensuremath{\mathrm{km}}}\xspace}
\newcommand{\Msun}{{\ensuremath{\mathrm{M}_{\odot}}}\xspace}
\newcommand{\Sec}{{\ensuremath{\mathrm{s}}}\xspace}
\newcommand{\erg}{{\ensuremath{\mathrm{erg}}}\xspace}

\newcommand{\kms}{{\ensuremath{\km\,\Sec^{-1}}}\xspace}

\newcommand{\gcc}{{\ensuremath{\g\,\cm^{-3}}}\xspace}

\newcommand{\kB}{{\ensuremath{k_{\mathrm{B}}}}\xspace}
\newcommand{\R}{{\ensuremath{\mathcal{R}}}\xspace}
\newcommand{\ms}{{\ensuremath{\mathrm{ms}}}\xspace}

\newcommand{\Junit}{{\ensuremath{\erg\,\Sec}}\xspace}
\newcommand{\Iunit}{{\ensuremath{\g\,\cm^2}}\xspace}
\newcommand{\wunit}{{\ensuremath{\mathrm{rad}\,\Sec^{-1}}}\xspace}
\newcommand{\G}{{\ensuremath{\mathrm{G}}}\xspace}
\newcommand{\Sunit}{{\ensuremath{\kB/\mathrm{baryon}}}\xspace}

\newcommand{\Msuns}{{\ensuremath{\Msun\,\Sec^{-1}}}\xspace}
\newcommand{\gps}{{\ensuremath{\g\,\Sec^{-1}}}\xspace}

\newcommand{\lSect}[1]{{\label{sec:#1}}}
\newcommand{\lFig}[1]{{\label{fig:#1}}}
\newcommand{\lEq}[1]{{\label{eq:#1}}}
\newcommand{\lTab}[1]{{\label{tab:#1}}}

\newcommand{\Tabff}[1]{{\ref{tab:#1}}}
\newcommand{\Tab}[1]{{Table~\Tabff{#1}}}

\newcommand{\TabRange}[2]{{Tables~\Tabff{#1} - \Tabff{#2}}}
\newcommand{\TabTwo}[2]{{Tables~\Tabff{#1} and \Tabff{#2}}}
\newcommand{\pan}[1]{{\textit{#1}}}

\newcommand{\FIGFF}[2]{{\ref{fig:#2}\pan{#1}}}

\newcommand{\FIG}[2]{{Fig.~\FIGFF{#1}{#2}}}
\newcommand{\Fig}[1]{{\FIG{}{#1}}}
\newcommand{\FigTwo}[2]{{\FIGS{}{#1} and \FIGFF{}{#2}}}
\newcommand{\FIGS}[2]{{Figs.~\FIGFF{#1}{#2}}}

\newcommand{\Sectff}[1]{{\ref{sec:#1}}}
\newcommand{\Sect}[1]{{\S\Sectff{#1}}}

\newcommand{\Appendices}[1]{{Appendices~\Sectff{#1}}}

\newcommand{\Eqref}[1]{{\ref{eq:#1}}}
\newcommand{\Eqff}[1]{{(\Eqref{#1})}}

\newcommand{\Eq}[1]{{Eq.~\Eqff{#1}}}
\newcommand{\Eqs}[1]{{Eqs.~\Eqff{#1}}}

\newcommand{\isofont}[1]{{\mathrm{#1}}}
\newcommand{\isomass}[1]{{\ensuremath{\isofont{^{#1}}}}}
\newcommand{\isocharge}[1]{{\ensuremath{\isofont{_{#1}}}}}
\newcommand{\isotope}[3]{{\ensuremath{\isocharge{#1}\isomass{#2}\isofont{#3}}}}

\newcommand{\I}[2]{{\isotope{}{#1}{#2}}}
\newcommand{\El}[1]{{\I{}{#1}}}

\newcommand{\Ep}[1]{{\ensuremath{10^{#1}}}}
\newcommand{\E}[1]{{\ensuremath{\powersep\Ep{#1}}}}
\newcommand{\EE}[2]{{\ensuremath{\powersep\Ep{#1#2}}}}

\newcommand{\Gone}{{\ensuremath{\Gamma\!_1}}\xspace}
\newcommand{\Gam}[1]{{\ensuremath{\Gamma\!_{#1}}}}
\newcommand{\gad}{{\ensuremath{\gamma\!_{\mathrm{ad}}}}\xspace}
\newcommand{\Nmu}{{\ensuremath{\nabla\!\!_{\mu}}}\xspace}
\newcommand{\Nad}{{\ensuremath{\nabla\!_{\mathrm{ad}}}}\xspace}

\newcommand{\NNmu}{{\ensuremath{N^2_{\mu}}}\xspace}
\newcommand{\NNcomp}{{\ensuremath{N^2_{\mathrm{comp}}}}\xspace}
\newcommand{\NNT}{{\ensuremath{N^2_{T}}}\xspace}
\newcommand{\NN}{{\ensuremath{N^2}}\xspace}

\newcommand{\Qc}{{\ensuremath{q_0}}\xspace}
\newcommand{\Qt}{{\ensuremath{q_1}}\xspace}

\newcommand{\Qm}{{\ensuremath{q_{\mathrm{min}}}}\xspace}

\newcommand{\Nue}{{\ensuremath{\nu_{\mathrm{e}}}}\xspace}
\newcommand{\Nure}{{\ensuremath{\nu_{\mathrm{re}}}}\xspace}
\newcommand{\Nuec}{{\ensuremath{\nu_{\mathrm{e0}}}}\xspace}
\newcommand{\Nuet}{{\ensuremath{\nu_{\mathrm{e1}}}}\xspace}
\newcommand{\Etae}{{\ensuremath{\eta_{\mathrm{e}}}}\xspace}
\newcommand{\De}{{\ensuremath{D_{\mathrm{e}}}}\xspace}
\newcommand{\Dec}{{\ensuremath{D_{\mathrm{e0}}}}\xspace}
\newcommand{\Det}{{\ensuremath{D_{\mathrm{e1}}}}\xspace}
\newcommand{\Nusc}{{\ensuremath{\nu_{\mathrm{sc}}}}\xspace}

\newcommand{\HP}{{\ensuremath{H_{\mathrm{P}}}}\xspace}
\newcommand{\Vconv}{{\ensuremath{v_{\mathrm{conv}}}}\xspace}
\newcommand{\cP}{{\ensuremath{c_{\mathrm{P}}}}\xspace}

\newcommand{\Br}{{\ensuremath{B_r}}\xspace}
\newcommand{\Bphi}{{\ensuremath{B_{\phi}}}\xspace}
\newcommand{\Brf}{{\ensuremath{B_r^4}}\xspace}
\newcommand{\Bphif}{{\ensuremath{B_{\phi}^4}}\xspace}
\newcommand{\Oc}{{\ensuremath{\Omega}}\xspace}
\newcommand{\Rhoc}{{\ensuremath{\rho_{\mathrm{c}}}}\xspace}
\newcommand{\BB}{{\ensuremath{\Bphi\Br}}\xspace}
\newcommand{\BoB}{{\ensuremath{\Br/\Bphi}}\xspace}
\newcommand{\Mgrav}{\ensuremath{M_{\mathrm{grav}}}\xspace}
\newcommand{\vwind}{\ensuremath{v_{\mathrm{wind}}}\xspace}
\newcommand{\rA}{\ensuremath{r_{\!\!\mathrm{A}}}\xspace}
\newcommand{\Mdot}{\ensuremath{\dot{M}}\xspace}

\slugcomment{Version \today}

\shorttitle{Magnetic Rotating Massive Stars}
\shortauthors{Heger, Woosley, \& Spruit}

\begin{document}

\title{Presupernova Evolution of Differentially Rotating\\
Massive Stars Including Magnetic Fields}

\author{A.~Heger}
\affil{Theoretical Astrophysics Group T6, MS B227, Los Alamos National Laboratory, Los Alamos, NM 87545; and\\
Enrico Fermi Institute, The University of Chicago, 5740 S.\ Ellis Ave, Chicago, IL 60637}
\email{alex@ucolick.org}
\author{S.~E.~Woosley}
\affil{Department of Astronomy and Astrophysics, University of California,
    Santa Cruz, CA 95064}
\email{woosley@ucolick.org}

\and
\author{H.~C.~Spruit}
\affil{Max Planck Institute for Astrophysics, Box 1317, 85741 Garching, Germany}
\email{hspruit@mpa-garching.mpg.de}

\begin{abstract}
As a massive star evolves through multiple stages of nuclear burning
on its way to becoming a supernova, a complex, differentially rotating
structure is set up.  Angular momentum is transported by a variety of
classic instabilities, and also by magnetic torques from fields
generated by the differential rotation.  We present the first stellar
evolution calculations to follow the evolution of rotating massive
stars including, at least approximately, all these effects, magnetic
and non-magnetic, from the zero-age main sequence until the onset of
iron-core collapse.  The evolution and action of the magnetic fields
is as described by \citet{Spr02} and a range of uncertain parameters
is explored.  In general, we find that magnetic torques decrease the
final rotation rate of the collapsing iron core by about a factor of
30 to 50 when compared with the non-magnetic counterparts.  Angular
momentum in that part of the presupernova star destined to become a
neutron star is an increasing function of main sequence mass.  That is,
pulsars derived from more massive stars will rotate faster and
rotation will play a more dominant role in the star's explosion.  The
final angular momentum of the core is determined - to within a factor
of two - by the time the star ignites carbon burning.  For the lighter
stars studied, around $15\,\Msun$, we predict pulsar periods at birth
near $15\,\ms$, though a factor of two range is easily tolerated by the 
uncertainties.  Several mechanisms for additional braking in a young
neutron star, especially by fall back, are also explored.
\end{abstract}


\section{Introduction}
\lSect{intro}

Massive stars are known to be rapid rotators with typical equatorial
velocities around 200\,\kms or more during hydrogen burning
\citep{Fuk82}.  It has long been known that this much rotation could
have a significant influence on the evolution of the star, both on the
main sequence and during in its later stages of evolution
\citep[e.g.,][]{ES76,ES78,HLW00,MM01,HMM04}.  For example, on the main
sequence, surface abundance patterns are observed
\citep[e.g.,][]{GL92,Her94,vra00,Ven99} that are most naturally
explained by rotationally-induced mixing in the stellar interior
\citep[e.g.,][]{Ven99,HL00,MM00}.

Numerical studies of of the later evolutionary stages of rotating
massive stars initially found that the stellar core would reach
critical (Keplerian) rotation by the time carbon ignited in the core
\citep[e.g.,][]{KMT70,ES76}, especially when angular momentum
transport is neglected.  Thus one might encounter triaxial deformation
early on, with continued evolution hovering near instability.  Modern
stellar models that include more instabilities capable of transporting
angular momentum find that considerable angular momentum is lost from
the core during hydrogen and helium burning \citep{ES78}, especially
when the stabilizing effect of composition gradients is reduced
\citep{Pi89,HLW00,MM01,HMM04}.  Such stars now evolve to iron core
collapse, but still form very rapidly rotating ($\sim1\,\ms$) neutron
stars.

However, with the exception of \citet{SP98} and \citet{MM04}, all
studies of massive stellar evolution to date have ignored what is
probably a major effect, the torques exerted in differentially
rotating regions by the magnetic fields that thread them.  This
omission has not been because such torques were thought to be
unimportant, but because of the complexity and uncertainty of carrying
out even one-dimensional calculations that included them.  This
uncertainty, in turn, related to the absence of a credible physical
theory that would even qualitatively describe the separate evolution
of the radial and poloidal components of the field, both of which are
necessary to calculate the torque,
\begin{equation}
S = \frac{B_r B_{\phi}}{4 \pi}
\end{equation} 

\citet{SP98} estimated these torques assuming that the poloidal field
results from differential winding and that the radial field would, due
to the action of unspecified instabilities, be comparable.  Particular
attention was paid to the region of large shear that separates the
core of helium and heavy elements from the very slowly rotating
hydrogen envelope of a red supergiant star.  This assumption, $\Br
\sim \Bphi$, probably overestimated the actual $\Br$ by orders of
magnitude and the rotation rates that \citet{SP98} estimated for
young pulsars were thus too slow (see also \citealt{Liv98}).

More recently, \citet{Spr02} has provided improved estimates for $\Br$
and $\Bphi$ based upon a dynamo process that takes into account the
effect of stable stratifications.  Here we investigate the effect of
this prescription in a stellar evolution code used to study the
complete evolution of massive stars.  Particular attention is given to
the rotation rate it implies for young neutron stars.  In \Sect{phys},
we describe the physical and numerical modeling.  In \Sect{result},
the results for $12$, $15$, $20$, $25$ and $35\,\Msun$ stars are
presented and, in \Sect{obs}, compared with observational data.
Typical neutron star rotation rates, at birth, are $\sim$ $15\,\ms$.
In \Sect{brake} we discuss other possible ways that such rapidly
rotating stars might be slowed during the first days of their
evolution by magnetic winds and fall back.  The surface abundance
changes due to rotationally-induced mixing are discussed in
\Sect{abundances} and in \Sect{conclude} we give our conclusions.

\section{Implementation of Magnetic Braking}
\lSect{phys}

The treatment of magnetic torques given by \citet{Spr02} was
implemented in a version of the implicit hydrodynamics stellar
evolution code \citep[KEPLER,][]{WZW78} that already included angular
momentum transport and mixing due to a number of non-magnetic
processes.  See \citet{HLW00} for a discussion of these non-magnetic
instabilities and \citet{Spr99,Spr02} for a detailed description of
the dynamo mechanism.  Here we discuss only the implementation of and
results from that physical model.  

Both angular momentum transport and chemical mixing are calculated by
solving the time-dependent diffusion equation \citep{HLW00}. It
is assumed that the dynamo adjusts and reaches equilibrium field
strength on a time-scale that is short compared to the evolution
time-scales of the star, i.e., we neglect the time required to reach
the steady state field strength and distribution described by
\citet{Spr02} and only apply their equilibrium values.  This should be
a reasonable approximation in all but the most advanced burning stages
that occur when the angular momentum distribution has already been
frozen in.  The validity of this assumption is revisited in
\Sect{conclude}

\subsection{The dynamo process in radiative regions}

\citet{Spr02} discusses two limiting cases: stabilization by
composition gradients (Spruit's Case 0), and stabilization by
superadiabatic temperature gradients (Spruit's Case 1).  Realistic
stellar models exhibit both kinds of gradients at same time, so the
intermediate case must also be included (see also \citealt{MM04}).

The effective viscosity, $\Nue$, for the radial transport of angular
momentum results from azimuthal stress due to the field generated by
the dynamo, $S=\Br\Bphi/4\pi$.  Its value is given by
\begin{equation} 
\Nue \equiv S/\rho q \Omega \;.
\lEq{nu}
\end{equation}
See also eqs.\ 34-40 of \citet{Spr02}.  We also implement the
``effective diffusivity'', $\De\equiv\Etae$, given in his eqs.\
41-43.  Here $\Omega$ is the angular velocity,
$q=\mathrm{d}\ln\Omega/\mathrm{d}\ln r$ is the shear, $r$ is the
radius, $\rho$ is the density, and $S$ is the stress.

For application outside the regime of ``ideal gas with radiation''
where the gradient of the mean molecular weight determines the
stabilization, we replace \NNmu by a more general formulation
(``$N^2_\mathrm{\!comp}$''), but do not change nomenclature. This way
the discussion in \citet{Spr02} can be followed without further
revision.  Generally, the Bruntv\"as\"al\"a frequency is
\begin{equation}
  \NN=\frac{g}{\HP}\left(\frac{1}{\Gone}
    -\frac{\mathrm{d} \ln \rho}{\mathrm{d} \ln P}\right)\;,
\end{equation}
where $g=-G M(r)/r^2$ is the local gravitational acceleration, $P$ is
the pressure, $\HP=-\mathrm{d} r / \mathrm{d}\ln P$ is the pressure
scale height, and the derivatives are total derivatives in the star.
The compositional contribution to the Bruntv\"as\"al\"a frequency is
given by
\begin{equation}
\NNmu=\frac{g}{\HP}\left(\alpha 
 - \delta \frac{\mathrm{d} \ln T}{\mathrm{d} \ln P} 
 - \frac{\mathrm{d} \ln \rho}{\mathrm{d} \ln P}\right)
\;,
\end{equation}
where $T$ is the temperature.  The thermodynamic quantities $\alpha$,
$\delta$, and $\Gone$ obey their common definitions (e.g.,
\citealt{KW90}).  Again, the derivatives are the actual gradients in the
star, not thermodynamic derivatives.  The thermal contribution to
the Bruntv\"as\"al\"a frequency is then given by
\begin{equation}
\NNT=\NN-\NNmu
=\frac{g}{\HP}\left(\frac{1}{\Gone} - \alpha 
 + \delta \frac{\mathrm{d} \ln T}{\mathrm{d} \ln P}
 \right)
\;.
\end{equation}
As stated above, we use the symbol $\NNmu$ to indicate (the square of)
the Bruntv\"as\"al\"a frequency due to the composition gradient for a general
equation of state ($\equiv\NNcomp$), not just due to changes of mean
molecular weight, $\mu$.  \Appendices{adi} and \Sectff{brunt} give a
derivation of the general expressions.  This formulation is identical
with the stability analysis for the different regimes (next section)
as used in KEPLER, and therefore necessary for consistency in regimes
where the assumption of an ideal gas is a poor approximation.

\subsection{The dynamo in semiconvective and thermohaline regions}

In addition to the radiative regime, \textbf{I}, where both
composition gradients and temperature gradients are stabilizing, other
regimes have to be considered in which ``secular'' mixing processes
operate. These are secular as opposed to processes like convection
that operate on a hydrodynamic time scale.  Two of these are 
semiconvection (\textbf{II}; where a stabilizing composition gradient
dominates a destabilizing temperature gradient) and thermohaline
convection, also called the salt-finger instability
\cite[\textbf{III}; where a stabilizing temperature gradient dominates
a destabilizing composition gradient; e.g.,][their Figure~1 for an
overview]{Bra97,HLW00}.  The final regime of hydrodynamic
instability in non-rotating stars is that of Ledoux convection
(\citealt{Led58}; Regime~\textbf{IV}; wehere the destabilizing
temperature gradient is stronger than any stabilization by the
composition gradient).  In the Ledoux regime, effective viscosities
and diffusivities are given by mixing length theory \citep{HLW00}.  A
very approximate treatment of Regimes {\bf II} and {\bf III} is given
in the next two sections.

\subsubsection{The semiconvective case}
\lSect{sc}

In semiconvective regions (Regime~\textbf{II}), the square of the
thermal buoyancy frequency is negative and thus the limiting Case
\textbf1 would be convection.  Instead of equation (34) of
\citet{Spr02}, we first compute an ``dynamo'' effective viscosity by
\begin{equation}
\Nure=\Nuec f(q)\;,
\end{equation}
with \Nuec from equation (35) of \citet{Spr02} and $\Qm=\Qc$
(equations 37-39 of \citealt{Spr02}).

Adopting a model for semiconvection in the non-linear regime similar
to \citet{Spr92}, we assume the radiative flux is transported by
convective motions in the semiconvective layers.  For the sake of
simplicity, we compute an effective turbulent viscosity, \Nusc, from
the velocities according to mixing-length theory \citep{SH58} assuming
Schwarzschild convection (i.e., when the $\mu$ gradient is neglected),
\begin{equation}
\Vconv \equiv 
\left(\frac{g \delta \HP L}{64 \pi \rho \cP T r^2}\right)^{\!1/3}
\end{equation}
\citep[e.g., as derived from equations 7.6 and 7.7 of][]{KW90}
and an assumed characteristic length scale of \HP by
\begin{equation}
\Nusc \equiv \frac{1}{3} \HP  \Vconv , 
\end{equation}
\citep[however, see][]{Spr92}, and \cP is the specific heat capacity
at constant pressure.

Finally, we assume that the actual viscosity is somewhere between
that of the case of mere semiconvection and that of the dynamo
dominated by composition stratification and take the geometric
mean between the two viscosities,
\begin{equation}
\Nue \equiv \sqrt{\Nure \Nusc} \; \lEq{nusc}.
\end{equation}
The effective turbulent diffusion coefficient, $\De$, is computed in an
analogous way, \Nure, i.e., $\De=\Dec f(q)$, $\Qm=\Qc$.  Due to the
properties of the layers semiconvection model \citep{Spr92} --
negligible diffusivity in typical stellar environments --, we do not
compute a geometric mean with the semiconvective diffusivity similar
to \Eq{nusc}.  Instead we just add it to the semiconvective diffusion
coefficient computed in KEPLER.  Typically it is negligible.

\subsubsection{The thermohaline case}
\lSect{th}

In this regime (\textbf{III}) the compositional buoyancy frequency
becomes negative and the limiting Case \textbf0 would be Rayleigh-Taylor
unstable.  We assume the limiting Case \textbf1 of \citet{Spr02} and
compute $\Nue=\Nuet$, $\De=\Det$, and $\Qm=\Qt$, analogous to the
semiconvective case.

In massive stars, thermohaline convection typically occurs following
radiative shell burning in the fuel ``trace'' left behind by an
receding central convective burning phase. The higher temperature of
shell burning can sometimes create heavier ashes than core burning and
the resulting compositional structure is unstable. The contraction
phase after central helium burning is the most important domain in
massive stars.

Generally, the motions associated with the growth of this instability
(``salt fingers'') are very slow.  Thus the interaction with the
dynamo is much more restricted than in the case of semiconvection.
Therefore we do \emph{not} employ a interpolation between the dynamo
effective radial viscosity or diffusivity as is done in \Sect{sc}.

\subsection{Evaluating the  magnetic field strength}

The most interstesting result of the dynamo model is the magnetic
field strength.  Combining equations (21), (23), (35) and (36) of
\citet{Spr02}, one has
\begin{equation}
  \Nue=r^2\Omega\left(\frac{\Br}{\Bphi}\right)^{\!2}\;.
\end{equation}
Using his equation (28), the components of the field are:
\begin{equation}
  \Brf=16\pi^2\rho^2\Nue^3q^2\Omega r^{-2}
\end{equation}
\begin{equation}
  \Bphif=16\pi^2\rho^2\Nue q^2\Omega^3 r^2
\;.
\end{equation}

\section{Results Including Magnetic Torques}
\lSect{result}

\subsection{Implementation}

Based upon the assumptions in \Sect{phys}, the full evolution of stars
of $12$, $15$, $20$, $25$, and $35\,\Msun$ of solar metallicity was
calculated.  For comparison, equivalent models were also calculated
without magnetic fields for three of these stars.  The initial models
and input physics were the same as in \citet{rau01}, but the total
angular momentum of the star was chsen such that a typical initial
equatorial rotation velocity of $\sim 200\,\kms$ \citep{Fuk82} was
reached on the zero-age main sequence (ZAMS; see \citealt{HLW00}).
Time-dependent mixing and angular momentum transport were followed as
discussed by \citet{HLW00} with the turbulent viscosities and
diffusivities from the dynamo model added to those from the model for
the hydrodynamic instabilities.  For the time being, possible
interactions between the two are neglected (though see
\citealt{MM04}).  Our standard case (\textbf{B}) implements the
description as outlined in the previous sections.

\subsection{Presupernova Models and Pulsar Rotation Rates}

\TabRange{15evo}{Xsurf} and \Fig{bfig} summarize our principal
results.  As also noted by \citet{MM04}, the inclusion of magnetic
torques with a magnitude given by \citet{Spr02}, greatly decreases the
differential rotation of the star leading to more slowly rotating
cores.

For illustration, we consider in some detail the evolution of the
$15\,\Msun$ model with standard parameter settings.  \Tab{15evo} shows
that, over the course of the entire evolution, the total angular
momentum enclosed by fiducial spheres of mass coordinate $1.5$, $2.5$,
and $3.5\,\Msun$ decreases in the magnetic models by a factor of 100
to 200, far more than in the non-magnetic comparison model.  In both
cases most of the angular momentum transport occurs prior to carbon
ignition (\FigTwo{jfig15}{jfig25}).  The greatest fractional decrease
occurs during the transition from hydrogen depletion to helium
ignition (\Tab{15evo}, \Fig{jfig15}) as the star adjusts to its new
red giant structure.  During this transition, the central density goes
up by a factor of about 100 (from $11\,\gcc$ to $1400\,\gcc$) with an
accompanying increase in differential rotation and shear.  Though the
angular momentum evolves little after carbon burning, modified only by
local shells of convection, there is a factor of two change after
carbon ignition (central temperature = $5\E8\,\K$).

The magnetic torques are sufficiently strong to enforce rigid rotation
both on the main sequence and in the helium core during helium burning
for all masses considered (see also \citealt{MM04}).  The appreciable
braking that occurs during hydrogen and helium burning is therefore a
consequence of mass loss and evolving stellar structure, especially
the formation of a slowly rotating hydrogen envelope around a rapidly
rotating helium core.  Typically the more massive stars spend a shorter
time during the critical Kelvin-Helmholtz contraction between hydrogen
depletion and helium ignition and also have a shorter helium burning
lifetime.  Hence, even though they actually have a little \emph{less}
angular momentum at the fiducial mass points in \Tab{15evo} on the
main sequence, the more massive stars have \emph{more} angular
momentum in their cores at the end of helium burning.  This distinction
persists and the more massive stars give birth to more rapidly
rotating neutron stars.

Stars that include all the usual non-magnetic mechanisms for
rotational mixing and angular momentum transport \citep{HLW00}, but
which lack magnetic fields have a considerably different evolution and
end up with 30 to 50 times more angular momentum in that part of their
core destined to collapse to a neutron star (\TabTwo{15evo}{PSR},
\FigTwo{jfig15}{jfig25}).  Most notable is the lack of appreciable
braking of the helium core by the outer layers in the period between
hydrogen core depletion and helium ignition.  The post-carbon-ignition
braking that accounted for an additional factor of two in the magnetic
case is also much weaker.  In fact, the inner $1.5\,\Msun$ of the
$15\,\Msun$ model has only about five times less angular momentum at
core collapse than implied by strict conservation from the (rigidly
rotating) main sequence onwards.  As noted previously many times, such
a large amount of angular momentum would have important consequences
for the supernova explosion mechanism.  Rigidly rotating neutron stars
cannot have a period shorter than about $1\,\ms$.

\Tab{15B} gives the magnetic field components that exist in different
parts of the star during different evolutionary phases.  The magnetic
field is highly variable from location to location in the star and the
values given are representative, but not accurate to better than a
factor of a few (\Fig{bfig}), especially during the late stages.  As
expected, the toroidal field, which comes from differential winding is
orders of magnitude larger than the radial field generated by the
Tayler instability.  The product \BB scales very approximately as the
central density with the ratio $\BoB \sim \Ep{-3} - \Ep{-4}$.  Though
the formulae used here are expected to break down during core
collapse, a simple extrapolation to $\Ep{15}\,\gcc$ for the central
density of a neutron star suggests $\Bphi \sim \Ep{14}$ and $\Br \sim
\Ep{10}$.

\Tab{PSR} gives the expected pulsar rotation rates based upon a set of
standard assumptions.  Following \citet{Lat01}, it is assumed that all
neutron stars (with realistic masses) have a radius of $12\,\km$ and a
moment of inertia $0.35\,\Mgrav R^2$.  Adding the binding energy to
the gravitational mass, assumed in \Tab{PSR} to be $1.4\,\Msun$ gives
the baryonic mass of the core that collapsed to the neutron star,
$1.7\,\Msun$.  Taking the angular momentum inside $1.7\,\Msun$ from the
presupernova model, assuming conservation during collapse, and solving
for the period then gives the values in the table.

\subsection{Sensitivity To Uncertain Parameters}

It is expected that our results will be sensitive to several uncertain
factors including the effect of composition gradients, the efficiency
of the dynamo in generating magnetic field, and the initial angular
momentum of the star.  To some extent, these uncertainties are
ameliorated by the strong sensitivity of the expected magnetic torques
to rotation speed and differential shear (eq.~34--36 of
\citealt{Spr02}).  Composition gradients have very significant
influence on the resulting angular momentum transport by inhibiting
the action of the dynamo.  In particular, composition interfaces, as
resulting from the nuclear burning processes in combination with
convection, usually also show significant shear, but if the dynamo
action is suppressed by a steep change in mean molecular weight,
angular momentum can become ``trapped'' in a fashion similar to that
observed by \citet{HLW00}.  To illustrate the uncertainties of the
dynamo model, we studied several test cases where the stabilizing
effect of composition gradients is varied by multiplying \NNmu and
\NNT by 1/10.  We also explored multiplying the overall coefficient of
the torque in eq.~(1) by varying \BB by a factor of ten and multiplied
the initial rotational rate by 0.5 and 1.5 (\Tab{PSR}).

In three cases, the $20$ and $25\,\Msun$ with high rotation rate, and
the ``normal'' $35\,\Msun$ model, the star lost its hydrogen-rich
envelope due to mass loss either during ($25$ and $35\,\Msun$) or just
at the end ($20\,\Msun$) of it evolution.  The KEPLER code was not
able to smoothly run these models through the period where they lost
their last solar mass of envelope because violent pulsations were
encountered.  The last 0.5 to $1.0\,\Msun$ was thus removed abruptly
and the calculation continued for the resulting Wolf-Rayet star.
However, it is clear that removing the envelope removes an appreciable
braking torque on the helium core so it is reasonable that the
remnants of such stars rotate more rapidly.

\subsection{Variable Neutron Star Masses}

Since the iron core masses, silicon shell masses, and density profiles
all differ appreciably for presupernova stars in the 12 to $35\,\Msun$
range, it is not realistic to assume that they all make neutron stars
of the same mass.  Ideally, one would extract realistic masses from a
reliable, credible model for the explosion, but, unfortunately, such
models have not reached the point of making accurate predictions.
Still there are some general features of existing models that provide
some guidance.  As noted by \citet{WZW78}, the remnant mass cannot
typically be less than the mass of the neutronized iron core.
Otherwise one overproduces rare neutron-rich isotopes of the iron
group.  More important to the present discussion, successful
explosions, when they occur in modern codes, typically give ``mass
cuts'' in the vicinity of large entropy jumps.  The reason is that a
jump in the entropy corresponds to sudden decrease in the density.
This in turn implies a sudden fall off in accretion as the explosion
develops.  The lower density material is also more loosely bound by
gravity.  The largest jump in the vicinity of the iron core is
typically at the base of the oxygen burning shell and that has
sometimes been used to estimate neutron star masses \citep{Tim96}.
Recent studies by Thomas Janka (private communication) suggest a
numerical criterion, that the mass cut occurs where the entropy per
baryon equals $4\,\kB$.  In \Tab{PSR(m)}, instead of assuming a
constant baryonic mass of $1.7\,\Msun$ for the neutron star progenitor
(as in \Tab{PSR}), we take the value where the specific entropy is
$4\,\Sunit$.

As expected this prescription gives larger baryonic masses for the
remnant (``Baryon'' in the table).  The fraction of this mass that is
carried away by neutrinos is given by \citet{Lat01},
\begin{equation}
f = \frac{0.6 \beta}{1 - 0.5 \beta}
\end{equation}
where $\beta = G\Mgrav/Rc^2$.  When this is subtracted, one obtains
the gravitational masses, \Mgrav in \Tab{PSR(m)}.  Using these,
assuming once more a moment of inertia $I = 0.35 \Mgrav R^2$ and a
radius of $12\,\km$, and conserving angular momentum in the collapse
one obtains the period.  Because the angular momentum per unit mass
increases as one goes out in the star, using a larger mass for the
remnant increases its rotation rate.

The resulting values are similar to those in \Tab{PSR} for the same
``standard'' parameter settings, but show even more clearly the
tendency of larger stars to make more rapidly rotating pulsars.  They
also show an additional prediction of the model - that more massive
pulsars will rotate more rapidly.  The numbers in \Tab{PSR(m)} have
not been corrected for the fact that the typical neutrino, since it
last interacts at the edge of the neutron star, carries away more
angular momentum than the average for its equivalent mass.  Thus the
periods in \Tab{PSR(m)} can probably be multiplied by an additional
factor of $\sim1.2$ \citep{Jan04}.

\section{Comparison With Observed Pulsar Rotation Rates}
\lSect{obs}

\Tab{PSR-obs} gives the measured and estimated rotation rates and
angular momenta of several young pulsars at birth
\citep{Mus96,mar98,Kas94}.  To estimate the angular momenta in the
table, we have assumed, following \citet{Lat01}, that the moment of
inertia for ordinary neutron stars (not quark stars) is $I \approx
0.35 MR^2 \approx 1.4\EE45\,\Iunit$.  Here a constant fiducial
gravitational mass of $1.4\,\Msun$ is assumed.

Given the uncertainties in both our model and the extrapolation of
observed pulsar properties, the agreement, at least for the 12 -
$15\,\Msun$ models, is quite encouraging.  This true all the more so
when one realizes that the periods in \TabTwo{PSR}{PSR(m)} should be
multiplied by a factor of approximately $1.2 - 1.3$ to account for the
angular momentum carried away by neutrinos when the neutron star forms
\citep{Jan04}.  Then our standard, numerically typical supernova,
$15\,\Msun$, would produce a neutron star with angular momentum
$5.8\EE47\,\Junit$ (``PreSN'' entry in \Tab{15evo} divided by 1.3).  It
should also be noted that the most common core collapse events - by
number - will be in the $12$ to $20\,\Msun$ range because of the
declining initial mass function.  Moreover, stars above around $25$ -
$30\,\Msun$ may end up making black holes \citep{Fry99}.

What then is the observational situation?  \citet{mar98} have
discovered a pulsar in the Crab-like supernova remnant, N157B, in the
Large Magellanic Cloud, with a rotation rate of $16\,\ms$.  This is
probably an upper bound to its rotation rate at birth.  \citet{Gle96}
gives an estimated initial rotation rate for the Crab pulsar itself of
$19\,\ms$.  Estimating the initial rotation rate of the much larger
sample of pulsars, known to be rotating now much more slowly, is
fraught with uncertainty.  Nevertheless, PSR 0531+21 and PSR 1509-58
are also estimated to have been born with $\sim20\,\ms$ periods
\cite{Mus96} (though 0540-69 is estimated to have $P_0 \approx
39\,\ms$).  At the other extreme, \citet{Pav02} find for PKS1209-51/52
a rotation period of $424\,\ms$ and argue that the initial rate was
not much greater.  Similarly, \citet{Kra03} argue that the initial
rotation rate of PSR J0538+2817 was $139\,\ms$.

\section{Further Braking Immediately After the Explosion}
\lSect{brake}

\subsection{r-Mode Instability}

For some time it was thought that gravitational radiation induced by
the ``\textsl{r}-mode instability'' would rapidly brake young neutron
stars.  However, \citet{Arr03} found that the \textsl{r}-mode waves
saturate at amplitudes lower than obtained in previous numerical
calculations \citep{LTV01} that assumed an unrealistically large
driving force.  Much less gravitational radiation occurs and the
braking time for a rapidly rotating neutron star becomes millennia
rather than hours.  \citeauthor{Arr03} calculate that
\begin{equation}
\tau_{\mathrm spin \ down} = 2000 \ {\rm yr} \ (\frac{\alpha_e}{0.1})^{-1} \
(\frac{335 \ {\rm Hz}}{\nu})^{11},
\end{equation}
with $\alpha_e$, the maximum amplitude of the instability (for which
\citeauthor{Arr03} estimate 0.1 as an upper bound).  A $3\,\ms$ pulsar
will thus take about 2000 years to substantially brake and the more
slowly rotating neutron stars in \TabTwo{PSR}{PSR(m)} will take even
longer.

\subsection{Neutrino-Powered Magnetic Stellar Winds}

\citet{Woo04} and \cite{Tho04} have discussed the possibility that a
young, rapidly rotating neutron star will be braked by a magnetic
stellar wind.  This wind, powered by the neutrino emission of the
supernova, may also be an important site for $r$-process
nucleosynthesis \citep{Tho03}. 

During the first $10\,\Sec$ of its life a neutron star emits about
$20\,\%$ of its mass as neutrinos.  This powerful flux of neutrinos
passing through the proto-neutron star atmosphere drives mass loss and
the neutron star loses $\sim0.01\,\Msun$ \citep{DSW86,QW96,TBM01}.
Should magnetic fields enforce corotation on this wind out to a
radius of 10 stellar radii, appreciable angular momentum could be lost
from a $1.4\,\Msun$ neutron star (since $j \sim r^2 \omega$).  The
issue is thus one of magnetic field strength and its radial variation.

For a neutron star of mass $M$ and radius, $R_6$, in units of
$10\,\km$, the mass loss rate is approximately (more accurate
expressions are given by \citeauthor{QW96})
\begin{equation}
\dot M \ \approx \ \Ep{-3} \ \Msun \ {\rm s^{-1}}\
\left(\frac{L_{\nu,{\rm tot}}}{\Ep{53} \ \Junit}\right)^{5/3} \
{R_6}^{5/3} \ \left(\frac{1.4\,\Msun}{M}\right)^2.
\end{equation}
The field will cause corotation of this wind out to a radius
\citep{MS87} where
\begin{equation}
\rho \ (\vwind^2 \ + \  \omega^2 r^2) \ \approx \ \frac{B^2}{4
  \pi}.
\end{equation}
Calculations that include the effect of rotation on the
neutrino-powered wind have yet to be done, but one can estimate the
radial density variation from the one-dimensional models of
\citet{QW96}.  For a mass loss rate of $\Ep{-2}\,\Msuns$ they find a
density and radial speed at $100\,\km$ of $\Ep8\,\gcc$ and
$1000\,\kms$ respectively.  At this time, the proto-neutron star
radius is $30\,\km$ and these conditions persist for $\sim 1\,\Sec$.
Later, they find, for a mass loss rate of $\Ep{-5}\,\Msuns$ and a
radius of $10\,\km$, a density at $100\,\km$ of $\sim\Ep5\,\gcc$ and
velocity $2000\,\kms$.  This lasts $\sim 10\,\Sec$.  Unless $\omega$
is quite low, \vwind is not critical.  For $\omega \sim 1000\,\wunit$
the field required to hold $\Ep8\,\gcc$ in corotation at $100\,\km$ is
$\sim 3\EE14\,\G$; for $\Ep5\,\gcc$ it is $\Ep{13}\,\G$.

To appreciably brake such a rapidly rotating neutron star, assuming $B
\sim r^{-2}$, thus requires ordered surface fields of $\sim\Ep{15}$ -
$\Ep{16}\,\G$.  Similar conclusions have been reached independently by
\citet{Tho03}.  Such fields are characteristic of magnetars, but
probably not of ordinary neutron stars.  On the other hand if the
rotation rate were already slow, $\omega \sim 100\,\wunit$ ($P
\sim60\,\ms$), even a moderate field of $\Ep{14}\,\G$ could have an
appreciable effect.  It should also be kept in mind that the field
strength of a neutron star when it is $1$ - $10\,\Sec$ old could be
very different than thousands of years later when most measurements
have been made.

\subsection{Fallback and the Propeller Mechanism} 

After the first $1000\,\Sec$, the rate of accretion from fall back is
given \citep{MWH01} by
\begin{equation}
\Mdot \approx \ 10^{-7} \ t_5^{-5/3} \; \Msuns,
\lEq{mdot}
\end{equation}
with the time in units of $\Ep5\,\Sec$.  For the mass loss rate in
units of $\Ep{26}\,\gps$, we obtain
\begin{equation}
\Mdot_{26} \ \approx \ 2 \ t_5^{-5/3} \;\gps.
\lEq{mdot26}
\end{equation}
For a dipole field with magnetic moment $\mu_{30} = B_{12} {R_6}^3$,
with $B_{12}$ the surface field in units of 10$^{12}$ G, the infalling
matter will be halted by the field at the Alfven radius \citep{Alp01},
\begin{equation}
\rA \ = \ 6.8 \ \mu_{30}^{4/7} \ \Mdot_{26}^{-2/7} \ \km.
\lEq{rA}
\end{equation}
At that radius matter can be rotationally ejected, provided the
angular velocity there corresponding to co-rotation exceeds the
Keplerian orbital speed.  The ejected matter carries away angular
momentum and brakes the neutron star.  This is the propeller mechanism
(\citealt{IS75,Che89,LWB91,Alp01}, but see the critical discussion in
\citealt{RFS04}).

Obviously $\rA$ must exceed the neutron star radius ($10\,\km$ here)
if the field is to have any effect.  The above equations thus require
a strong field, $B > \Ep{12}\,\G$, and accretion rates characteristic
of an age of at least one day.  Additionally there is a critical
accretion rate, for a given field strength and rotation rate, above
which the co-rotation speed at the Alfven radius will be slower than
the Keplerian orbit speed.  In this case the matter will accrete
rather than be ejected.  Magnetic braking will thus be inefficient
until $\omega^2 \, \rA^3 \, > \, GM$, or
\begin{equation}
\dot M_{26} \ < \ 5.7\EE-4 \ \omega_3^{7/3} \ \mu_{30}^2,
\lEq{mdot26a}
\end{equation}
where $\omega_3$ is the angular velocity in thousands of radians per
second ($\omega_3 = 1$ implies a period of $2\pi\,\ms$).  This turns
out to be a very restrictive condition.  For a given $\mu_{30}$ and
$\omega_3$, \Eqs{mdot26} and \Eqff{mdot26a} give a time,
$t_{5,\mathrm{min}}$, when braking can begin.  The torque on the
neutron star from that point on will be
\begin{equation}
I \dot \omega \ = \ \mu^2/\rA^3  \ = 10^{60} 
\frac{\mu_{30}^2}{\rA^3}, 
\end{equation}
with $I$, the moment of inertia of the neutron star, approximately
$\Ep{45}$ \citep{Lat01}.
The integrated deceleration will be 
\begin{equation}
\Delta \omega \ = \ 250 \, \mu_{30}^{2/7} \, t_{5, \mathrm{min}}^{-3/7}.
\end{equation}

Putting it all together, a $1.4\,\Msun$ neutron star can be braked to
a much slower speed ($\Delta \omega \sim \omega$) by fall back if
$\mu_{30} > 78$, $43$, or $25$ for initial periods of $6$, $21$, or
$60\,\ms$, respectively.  If the surface field strength - for a dipole
configuration is less than $2\EE{13}\,\G$, braking by the propeller
mechanism will be negligible in most interesting situations.

However, we have so far ignored all non-magnetic forces save gravity
and centrifugal force.  A neutron star of age less than one day is in
a very special situation.  The accretion rate given by \Eq{mdot} is
vastly super-Eddington. This means that matter at the Alfven radius
will be braked by radiation as well as centrifugal force and the
likelihood of its ejection is greater.  \citet{FBH96} have considered
neutron star accretion at the rates relevant here and find that
neutrinos released very near the neutron star actually drive an
\emph{explosion} of the accreting matter.  Just how this would all
play out in a multi-dimensional calculation that includes magnetic
braking, rotation, and a declining accretion rate has yet to be
determined, but is worth some thought.

If the accreting matter is actually expelled under `propeller'
conditions, one might expect that the mechanism also inhibits the
accretion so the process might be self-limiting.  This traditional
view of the propeller mechanism may be misleading however.  There is a
range in conditions where accretion accompanied by \emph{spindown} is
expected to occur (\citealt{Sha77,Spr93}; for a recent discussion see
\citealt{RFS04}), as observed in the X-ray pulsars.  If fall back is
the way most neutron stars are slowed, one might expect a correlation
of pulsar period with the amount of mass that falls back and that
might increase with progenitor mass.

More massive stars may experience more fall back \citep{WW95} and make
slowly rotating neutron stars.  But, on the other hand, as
\TabTwo{PSR}{PSR(m)} show, neutron stars derived from more massive
stars are born rotating more rapidly.  There may be a mass around
$15\,\Msun$ or so where the two effects combine to give the slowest
rotation rate.  Interestingly, the Crab pulsar was probably once a
star of $\sim10\,\Msun$ \citep{Nom82} and may thus have experienced
very little fall back.  The matter that is ejected by the propeller
mechanism in more massive stars could contribute appreciably to the
explosion of the supernova and especially its mixing.

\subsection{Effect of a Primordial Field}

As noted above, the results agree with the initial rotation periods
found in Crab-like pulsars, but not with the long periods inferred for
pulsars like PSR J0538+2817.  Within the framework of our present
theory, slowly rotating pulsars must have formed from stars with
different or additional angular momentum transport mechanisms.
Another possibility is that their main sequence progenitors
started with a qualitatively different magnetic field configuration.
The dynamo process envisaged here assumes that the initial field of
the star was sufficiently weak, such that winding-up by differential
rotation produces the predominantly toroidal field in which the dynamo
process operates.  Any weak large scale field initially present is then
expelled by turbulent diffusion associated with the dynamo. 

Stars with strong initial fields, such as seen in the magnetic
A-stars, cannot have followed this path.  In these stars, the magnetic
field is likely to have eliminated any initial differential rotation
of the star on a short time scale \citep{Spr99}, after which it
relaxed to the observed stable configurations \citep{Bra04}.  In these
configurations the radial and azimuthal field components are of
similar magnitude.  To the extent that such magnetic fields also exist
in massive MS stars (where they are much harder to detect, see however
\citealt{Don02}), they are likely to have led to a stronger magnetic
coupling between core and envelope, and hence to more slowly rotating
pre-SN cores.

\section{Surface Abundances}
\lSect{abundances}

The abundances of certain key isotopes and elements on the surfaces of
main sequence stars and evolved supergiants are known to be
diagnostics of rotationally-induced mixing \citep{HL00,Mae00}.  In
particular, on the main sequence, one expects enhancements of \I4{He},
\I{13}C, \I{14}N, \I{17}O, and \I{23}{Na} and extra depletion of
\I{12}C, \I{15}N, \I{16}O, \I{18}O, \I{19}F, \El{Li}, \El{Be}, and
\El{B}.  To test these predictions and their sensitivity to assumed
magnetic torques in the present models, we calculated three additional
versions of our $15\,\Msun$ model in which a large nuclear reaction
network was carried in each zone.  The adopted nuclear physics and
network were essentially the same as in \citet{rau01}.  One model
assumed no rotation; a second assumed rotation without magnetic
torques; and a third, which included both rotation and magnetic
torques, was our standard model in \TabRange{15evo}{PSR(m)}.  The
initial rotation rate assumed in the magnetic and non-magnetic models
was the same.

The results (\Tab{Xsurf}) show the same sort of rotationally-induced
enhancements and depletions on the main sequence (defined by
half-hydrogen depletion at the center) as in \citet{HL00}.  Helium is
up by about half a percent and nitrogen is increased by two.  Carbon
is depleted by about $25\,\%$ and \I{15}N is depleted by a factor of
two.  The isotopes \I{13}C and \I{17}O are enhanced by factors of
three and $1.6$ respectively.  \El{Be} and \El{B} are depleted.

Most important to the current discussion, we see no dramatic
differences between the surface abundances calculated with and
without magnetic torques.  \citet{MM04} suggested that such
differences might be a discriminant, perhaps ruling out torques of the
magnitude suggested by \citet{Spr02}.

\section{Conclusions and Discussion}
\lSect{conclude}

Given the uncertainty that must accompany any first-principles
estimate of magnetic field generation in the interior of stars, the
rotation rates, we derive here compare quite favorably with those
inferred for a number of the more rapidly rotating pulsars.  Without
invoking any additional braking during or after the supernova
explosion, the most common pulsars, which come from stars of $10$ -
$15\,\Msun$, will have rotation rates around $10$ - $15\,\ms$
(\Tab{PSR}).  Reasonable variation in uncertain parameters could
easily increase this to $20\,\ms$.  Given that the torques in the
formalism of \citet{Spr02} depend upon the fourth power of the shear
between layers, this answer is robust to small changes in the overall
coupling efficiency and inhibition factors.  If we accept the premise
that only the fastest solitary pulsars represent their true birth
properties and the others have been slowed by fall back or other
processes occurring during the first few years of evolution, our
predictions are in agreement with observations.

If correct, this conclusion will have important implications, not only
for the evolution of pulsars, but for the supernova explosion
mechanism, gravitational wave generation, and for gamma-ray bursts
(GRBs).  For rotation rates more rapid than $5\,\ms$, centrifugal
forces are an important ingredient in modeling the collapse and the
kinetic energy available from rotation is more than the
$\Ep{51}\,\erg$ customarily attributed to supernovae.  At $10\,\ms$,
the energy is nearly negligible.

Given recent observational indications that supernovae accompany some
if not all GRBs of the ``long-soft'' variety (e.g., \citealt{Woo05}),
rotation is apparently important in at least {\sl some}
explosions. Otherwise no relativistic jet would form. Indeed, the only
possibilities are a pulsar with very strong magnetic field and
rotation rate $\sim1\,\ms$ (e.g., \citealt{Whe00}) and a rapidly
rotating black hole with an accretion disk \citep{Woo93,Mac99}. Both
require angular momenta considerably in excess of any model is
\Tab{PSR} except those that did not include magnetic torques.

We shall address this subject in a subsequent paper, but for now note
a possible important symmetry braking condition - the presence of a
red supergiant envelope. The most common supernovae - Type IIp - will
result from stars that spent an extended evolutionary period with a
rapidly rotating helium core inside a nearly stationary hydrogen
envelope. GRBs on the other hand come from Type Ic supernovae, the
explosions of stars that lost their envelopes either to winds or
binary companions. Whether a comprehensive model can be developed
within the current framework that accommodates both a slow rotation
rate for pulsars at birth and a rapid one for GRBs remains to be seen
but we are hopeful.

As noted in \Sect{phys}, the magnetic torques have been computed
assuming that the dynamo process is always close to steady state. The
time to reach this steady state scales as some small multiple of the
time for an Alfv\'en wave to travel around the star along the toroidal
field. This time is very short throughout most of the evolution, but
in the latest stages it eventually becomes longer than the time scale
on which the moment of inertia of the core changes.  After this, the
dynamo effectively stops tracking the changing conditions and the
field is frozen in (its components varying as $1/r^2$, for homologous
contraction). With the data in \TabTwo{15evo}{15B}, we can estimate
the effect this would have on the angular momentum loss from the
core. The Alfv\'en crossing time first exceeds the evolution time
around the end of carbon burning.  Evolving the field under frozen-in
conditions from this time on then turns out to produce field strengths
that do not differ greatly (less than a factor of 4) from those shown
in \Tab{15B}.  Since the angular momentum of the core does not change
by more than $5\,\%$ after carbon burning, the effect of using frozen
field conditions instead of a steady dynamo is therefore small.

\acknowledgments This research was supported, in part, by the NSF (AST
02-06111), NASA (NAG5-12036), and the DOE Program for Scientific
Discovery through Advanced Computing (SciDAC; DE-FC02-01ER41176).  AH
was also funded by the DOE under grant B341495 to the FLASH Center at
the University of Chicago, under DOE contract W-7405-ENG-36 to the Los
Alamos National Laboratory, and acknowledges supported by a Fermi
Fellowship of the Enrico Fermi Institute at The University of Chicago,
and the Alexander von Humboldt-Stiftung (FLF-1065004).

\clearpage

\begin{appendix}

\section{Thermodynamic derivatives}
\lSect{adi}

Instead of the formulations 
\begin{eqnarray}
\Gam1\equiv\gad=\frac{1}{\delta-\alpha\Nad},
\qquad
1-\frac{1}{\Gam2}\equiv\Nad=\frac{\R}{\delta\mu\cP}
\end{eqnarray}
for an ideal gas with radiation (\R is the gas constant; \citealt{KW90}),
the general expressions should be used for more general equations of
state, as to, e.g., include the effect of degeneracy which is
important in the late stages of stellar evolution of massive stars and
in low mass stars.  The thermodynamic derivatives at constant entropy
\begin{eqnarray}
\Gam1\equiv\gad=
\left(\frac{\mathrm{d}\ln P}{\mathrm{d}\ln\rho}\right)_{\!\!\mathrm{ad}}\\
1-\frac{1}{\Gam2}\equiv\Nad=
\left(\frac{\mathrm{d}\ln T}{\mathrm{d}\ln P}\right)_{\!\!\mathrm{ad}}\\
\Gam3-1=
\left(\frac{\mathrm{d}\ln T}{\mathrm{d}\ln\rho}\right)_{\!\!\mathrm{ad}}\\
\frac{\Gam1}{\Gam3-1}=\frac{\Gam2}{\Gam2-1}
\lEq{gam}
\end{eqnarray}
\citep{KW90} can be derived from $\mathrm{d}s=\mathrm{d}q/T$,
$\mathrm{d}q=\mathrm{d}u+P\mathrm{d}\left(1/\rho\right)=\mathrm{d}u-P/\rho^2\,\mathrm{d}\rho$,
and the total derivative on the specific internal energy
\begin{equation}
\mathrm{d}u=
\left(\frac{\partial u}{\partial\rho}\right)_{\!\!T}\mathrm{d}\rho+
\left(\frac{\partial u}{\partial T}\right)_{\!\!\rho}\mathrm{d} T
\end{equation}
assuming adiabatic changes, $\mathrm{d}s=0$:
\begin{equation}
0=\mathrm{d}s=\frac{\mathrm{d}q}{T}=\frac{1}{T}\left[
\left(\frac{\partial u}{\partial\rho}\right)_{\!\!T}
-\frac{P}{\rho^2}\right]\mathrm{d}\rho 
+ \frac{1}{T}\left(\frac{\partial u}{\partial T}\right)_{\!\!\rho}\mathrm{d}T
\lEq{ds}
\end{equation}
\citep{KW90}.  This can be transformed into
\begin{eqnarray}
\left(\frac{\mathrm{d}T}{\mathrm{d}\rho}\right)_{\!\!\mathrm{ad}}=
\left[
\frac{P}{\rho^2}-\left(\frac{\partial u}{\partial\rho}\right)_{\!\!T}
\right] \left/
\left(\frac{\partial u}{\partial T}\right)_{\!\!\rho}
\right.
\end{eqnarray}
and we readily obtain
\begin{eqnarray}
\Gam3-1=
\left(\frac{\mathrm{d}\ln T}{\mathrm{d}\ln\rho}\right)_{\!\!\mathrm{ad}}=
\frac{\rho}{T}
\left(\frac{\mathrm{d}T}{\mathrm{d}\rho}\right)_{\!\!\mathrm{ad}}=
\frac{\rho}{T}
\left[
\frac{P}{\rho^2}-\left(\frac{\partial u}{\partial\rho}\right)_{\!\!T}
\right] \left/
\left(\frac{\partial u}{\partial T}\right)_{\!\!\rho}
\right.
\end{eqnarray}
as a function of the partial derivatives of $u$ with respect to $T$
and $\rho$.  Using the equation of state for bubble without mixing or
composition exchange with its surrounding,
\begin{equation}
\frac{\mathrm{d}\rho}{\rho}=\alpha\frac{\mathrm{d}P}{P}-\delta\frac{\mathrm{d}T}{T}
\lEq{eos}
\end{equation}
where
\begin{eqnarray}
\alpha=\left(\frac{\partial\ln\rho}{\partial\ln P}\right)_{\!\!T}
=1\left/\left[\frac{\rho}{P}\left(\frac{\partial P}{\partial \rho}\right)_{\!\!T}
\right]\right.
\\
\delta=
-\left(\frac{\partial\ln\rho}{\partial\ln T}\right)_{\!\!P}
=\frac{T}{\rho}
\left(\frac{\partial P}{\partial T}\right)_{\!\!\rho}
\left/
\left(\frac{\partial P}{\partial \rho}\right)_{\!\!T}
\right.
\;.
\end{eqnarray}
and using the thermodynamic relation (e.g., \citealt{KW90})
\begin{eqnarray}
\left(\frac{\partial P}{\partial T}\right)_{\!\!\rho}=
\frac{P\delta}{T\alpha}
\end{eqnarray}
to transform the last term, we can write \Eq{ds} as
\begin{eqnarray}
0=\mathrm{d}s=
\left\{
  \frac{1}{T}
  \left[
     \left(\frac{\partial u}{\partial\rho}\right)_{\!\!T}
     - \frac{P}{\rho^2}
  \right] 
  - \frac{1}{\rho\delta}
    \left(\frac{\partial u}{\partial T}\right)_{\!\!\rho}
\right\}\mathrm{d}\rho
+
\frac{\alpha}{P\delta}
\left(\frac{\partial u}{\partial T}\right)_{\!\!\rho}
\mathrm{d}P
\end{eqnarray}
and obtain
\begin{eqnarray}
\Gam1=
\left(\frac{\mathrm{d}\ln P}{\mathrm{d}\ln\rho}\right)_{\!\!\mathrm{ad}}
=
\frac{\rho}{P}
\left(\frac{\mathrm{d}P}{\mathrm{d}\rho}\right)_{\!\!\mathrm{ad}}
=
\frac{\rho}{P}
\left\{
  \left[
    \frac{P}{\rho^2} 
    -
    \left(\frac{\partial u}{\partial\rho}\right)_{\!\!T}
  \right]
  \left[
    \left(\frac{\partial P}{\partial T}\right)_{\!\!\rho}
    \left/
      \left(\frac{\partial u}{\partial T}\right)_{\!\!\rho}
    \right.
  \right]
  +
  \left(\frac{\partial P}{\partial \rho}\right)_{\!\!T}
\right\}
\end{eqnarray}
as a function of the derivative of $u$ and $P$ with respect to $T$ and
$\rho$ only.  Using \Eq{gam} we can solve for $\Gam2$ employing the
relation for $\Gam3$:
\begin{eqnarray}
1-\frac{1}{\Gam2}
=
\frac{\Gam3-1}{\Gam1}
=
\left(\frac{\mathrm{d}\ln T}{\mathrm{d}\ln P}\right)_{\!\!\mathrm{ad}}
=
\frac{T}{P}
\left(\frac{\mathrm{d}T}{\mathrm{d}P}\right)_{\!\!\mathrm{ad}}
=
\left[
  P-\rho^2\left(\frac{\partial u}{\partial\rho}\right)_{\!\!T}
\right]
\left/
  \left[
    T\rho\,\Gam1\left(\frac{\partial u}{\partial T}\right)_{\!\!\rho}
  \right]
\right.
\;.
\end{eqnarray}

\section{Compositional Bruntv\"as\"al\"a frequency}
\lSect{brunt}

We derive the compositional fraction of the Bruntv\"as\"al\"a
frequency from the following consideration: we compare a change of
density if we move a fluid element to a new location and allow
adjusting it pressure and temperature to its surroundings.  When
moving to the new location, $P$ will be different by $\mathrm{d}P$ and
$T$ be different by $\mathrm{d}T$.  The density change we then obtain
from the equation of state, \Eq{eos}, for a displacement with such a
change in pressure thus is
\begin{equation}
\left(\frac{\mathrm{d}\ln\rho}{\mathrm{d}\ln P}\right)_{\!\!\mathrm{element}}=\alpha-\delta\left(\frac{\mathrm{d}\ln T}{\mathrm{d}\ln P}\right)_{\!\!\rho}
\end{equation}
The buoyancy is obtained by comparing with a density change in the
star over the same change in pressure, the remainder than is due to
compositional changes.
\begin{equation}
\left(\frac{\mathrm{d}\ln (\Delta\rho)}{\mathrm{d}\ln P}\right)_{\!\!\mathrm{comp}}
=
\alpha
-
\delta\frac{\mathrm{d}\ln T}{\mathrm{d}\ln P}
-
\frac{\mathrm{d}\ln \rho}{\mathrm{d}\ln P}
\lEq{buoy}
\end{equation}
For a compositionally stably stratified medium this quantity is
negative since pressure decreases outward.

As an example let us consider an ideal gas with radiation.  For the
surrounding, the equation of state is, as a function of $T$, $P$, and
mean molecular weight, $\mu$, given by 
\begin{equation}
\frac{\mathrm{d}\rho}{\rho}=\alpha\frac{\mathrm{d}P}{P}-\delta\frac{\mathrm{d}T}{T}+\varphi\frac{\mathrm{d}\mu}{\mu}
\end{equation}
with
\begin{equation}
\varphi=\left(\frac{\partial\ln\rho}{\partial\ln\mu}\right)_{\!\!T,P}
\;.
\end{equation}
Substituting this in for the last term of \Eq{buoy}, we obtain
\begin{equation}
\left(\frac{\mathrm{d}\ln (\Delta\rho)}{\mathrm{d}\ln P}\right)_{\!\!\mathrm{comp}}
=
-\varphi
\frac{\mathrm{d}\ln\mu}{\mathrm{d}\ln P}
\equiv
-\varphi\Nmu\;,
\end{equation}
and the usual expression for buoyancy in an ideal gas with radiation is recovered.

For a displaced blob of material the square of the oscillation
``angular'' frequency ($\nu=1/\mathrm{period}$), the Bruntv\"as\"al\"a
frequency, can be obtained by multiplication with $g/\HP$, analogous to
the harmonic oscillator, here for the case of isothermal displacements
that only consider stabilization by composition gradients:
\begin{equation}
\NNcomp=\frac{g}{\HP}
\left(
\alpha
-
\delta\frac{\mathrm{d}\ln T}{\mathrm{d}\ln P}
-
\frac{\mathrm{d}\ln \rho}{\mathrm{d}\ln P}
\right)
\end{equation}

\end{appendix}

\onecolumn


\begin{table}
\caption{Evolution of Angular Momentum at Fiducial Mass Coordinates for $15\,\Msun$ star
\lTab{15evo}}
\smallskip
\begin{tabular}{lcccccc}
\hline\hline\noalign{\smallskip}
& \multicolumn{3}{c}{\hrulefill magnetic star\hrulefill} 
& \multicolumn{3}{c}{\hrulefill non-magnetic star\hrulefill} 
\\ 
& J(1.5) & J(2.5) & J(3.5) 
& J(1.5) & J(2.5) & J(3.5)   \\
\hline\noalign{\smallskip}
ZAMS        & 1.75\E{50} & 4.20\E{50} & 7.62\E{50} & 2.30\E{50} & 5.53\E{50} & 1.00\E{51} \\
H-burn$^a$  & 1.31\E{50} & 3.19\E{50} & 5.83\E{50} & 1.51\E{50} & 3.68\E{50} & 6.72\E{50} \\ 
H-dep$^b$   & 5.02\E{49} & 1.26\E{50} & 2.37\E{50} & 1.36\E{50} & 3.41\E{50} & 6.37\E{50} \\ 
He-ign$^c$  & 4.25\E{48} & 1.21\E{49} & 2.57\E{49} & 1.16\E{50} & 2.98\E{50} & 4.87\E{50} \\
He-burn$^d$ & 2.85\E{48} & 7.84\E{48} & 1.83\E{49} & 7.06\E{49} & 1.85\E{50} & 3.86\E{50} \\
He-dep$^e$  & 2.23\E{48} & 5.95\E{48} & 1.21\E{49} & 4.72\E{49} & 1.26\E{50} & 2.52\E{50} \\
C-ign$^f$   & 1.88\E{48} & 5.52\E{48} & 1.12\E{49} & 4.69\E{49} & 1.26\E{50} & 2.46\E{50} \\
C-dep$^g$   & 8.00\E{47} & 3.26\E{48} & 9.08\E{48} & 4.06\E{49} & 1.25\E{50} & 2.24\E{50} \\
O-dep$^h$   & 7.85\E{47} & 3.19\E{48} & 8.43\E{48} & 3.94\E{49} & 1.20\E{50} & 1.99\E{50} \\
Si-dep$^i$  & 7.76\E{47} & 3.05\E{48} & 7.23\E{48} & 3.75\E{49} & 1.16\E{50} & 1.95\E{50} \\
PreSN$^j$   & 7.55\E{47} & 2.59\E{48} & 7.31\E{48} & 3.59\E{49} & 1.09\E{50} & 1.94\E{50} \\
\hline
\end{tabular}
\smallskip\\
NOTE: 
$^a$40\,\% central hydrogen mass fraction; 
$^b$1\,\% hydrogen left in the core; 
$^c$1\,\% helium burnt;  
$^d$50\% central helium mass fraction; 
$^e$1\,\% helium left in the core; 
$^f$central temperature of 5\E8\,\K; 
$^g$central temperature of 1.2\E9\,\K; 
$^h$central oxygen mass fraction drops below 5\,\%;
$^i$central Si mass fraction drops below \Ep{-4}; 
$^j$infall velocity reaches 1000\,\kms. 
\end{table}

\begin{table}
\caption{Approximate Magnetic Field and Angular Velocity Evolution in a
a $15\,\Msun$ Star \lTab{15B}}
\smallskip
\begin{tabular}{lrrrrrr}
\hline\hline
                          & M$_{\rm samp}$ & $\rho_{\rm samp}$ & R$_{\rm samp}$
                          & \Bphi   & \Br     & \Oc     \\
\raisebox{1.5ex}[0pt]
{evolution stage}         & (\Msun) & (\gcc)  & (10$^9$ cm) & (\G)   & (\G)   &
                          (\wunit)  \\
\hline\noalign{\smallskip}
MS$^a$           \dotfill & 5.0     &  1.9  & 90 &  2\E{4} & 0.5 &
                            4\E{-5}  \\
TAMS$^b$         \dotfill & 3.5     &  2.7  & 67 &  3\E{4} & 1   & 
                            2\E{-5}  \\
He ignition$^c$  \dotfill & 3.5     &  0.90 & 52 & 5\E{3}   &  5  & 
                            4\E{-6}  \\
He ignition$^c$  \dotfill & 1.5     &  470  & 9.8  & 3\E{4}   & 20  & 
                            4\E{-5}  \\
He depletion$^d$ \dotfill & 3.5     & 150   & 14 &  2\E{4} & 3.5 & 
                            3\E{-5}  \\
C ignition$^e$   \dotfill & 1.5     &   1\E{4} & 3.5 & 6\E{5} & 250   & 
                             2\E{-4} \\
C depletion$^f$  \dotfill & 1.2     &   3\E{5} & 0.80 & 3\E{7} & 5\E{3} & 
                           1\E{-3}   \\
O depletion$^g$  \dotfill & 1.5     &   4\E{5} & 0.73 & 2\E{7} & 2\E{3} & 
                           1\E{-3}   \\
Si depletion$^h$ \dotfill & 1.5     &   2\E{6} & 0.44 & 5\E{7} & 5\E{3} & 
                           3\E{-3}   \\
pre-SN$^i$       \dotfill & 1.3     &   5\E{7} & 0.12 & 5\E{9} & 1\E{6} & 
                           5\E{-2}   \\
\hline\hline
                          & M$_{\rm samp}$ & \Rhoc & T$_{\rm c}$ 
                          & t$_{\rm death}$ & & \\
\raisebox{1.5ex}[0pt]
{evolution stage}         & (\Msun) & (\gcc)  & (10$^9$ K) & (sec) & & \\
\hline\noalign{\smallskip}
MS               \dotfill & 5.0     &  5.6  & 0.035 & 2.0\EE14 & & \\
TAMS             \dotfill & 3.5     &  11   & 0.045 & 6.4\EE13 & & \\
He ignition      \dotfill & 3.5     & 1400  & 0.159 & 6.0\EE13 & & \\
He ignition      \dotfill & 1.5     & 1400  & 0.159 & 6.0\EE13 & & \\
He depletion     \dotfill & 3.5     & 2700  & 0.255 & 1.4\EE12 & & \\
C ignition       \dotfill & 1.5     & 3.8\E{4} & 0.50 & 2.4\EE11 & &  \\
C depletion      \dotfill & 1.2     & 7.0\E{6} & 1.20 & 3.4\E8 & & \\
O depletion      \dotfill & 1.5     & 1.0\E{7} & 2.20 & 1.1\E7 & & \\
Si depletion     \dotfill & 1.5     & 4.8\E{7} & 3.76 & 8.3\E4 & & \\
pre-SN           \dotfill & 1.3     & 8.7\E{9} & 6.84 & 0.5 & &  \\
\hline
\end{tabular}
\smallskip\\
NOTE: \\
$^a$40\,\% central hydrogen mass fraction; \\
$^b$1\,\% hydrogen left in the core; \\
$^c$1\,\% helium burnt;  \\
$^d$1\,\% helium left in the core; \\
$^e$central temperature of 5\E8\,\K; \\
$^f$central temperature of 1.2\E9\,\K; \\
$^g$central oxygen mass fraction drops below 5\,\%; \\
$^h$central Si mass fraction drops below \Ep{-4}; \\
$^i$infall velocity reaches 1000\,\kms. 
\end{table}

\begin{table}
\caption{Pulsar Rotation Rate Dependence on Dynamo Model Parameters$^a$
\lTab{PSR}}
\smallskip
\begin{tabular}{lrrrrrrrrrrr}
\hline\hline\noalign{\smallskip}
& 
& \multicolumn{2}{c}{\hrulefill\,\NNmu\hrulefill}
& \multicolumn{2}{c}{\hrulefill\,\NNT\hrulefill}
& \multicolumn{2}{c}{\hrulefill\,\BB\hrulefill}
& \multicolumn{2}{c}{\hrulefill\,$\Omega_\mathrm{ZAMS}$\hrulefill}
\\
\raisebox{1.5ex}[0pt]{initial} 
& \raisebox{1.5ex}[0pt]{``std.''} 
& 0.1
&  10
& 0.1
&  10
& 0.1
&  10
& 0.5
& 1.5
&\multicolumn{2}{c}{B=0}
\\
\raisebox{1.5ex}[0pt]{mass} 
& \multicolumn{11}{c}{\dotfill period (\ms) \dotfill} \\
\hline\noalign{\smallskip}
12\,\Msun & 9.9 &    &     &     &     &     &    &     &     &      \\
15\,\Msun & 11  & 24 & 4.4 & 12  & 10  & 5.7 & 21 & 9.8 & 10      & 0.20 \\
20\,\Msun & 6.9 & 14 & 3.2 & 8.4 & 6.4 & 3.3 & 11 & 7.2 & 6.5$^b$ & 0.21 \\
25\,\Msun & 6.8 & 13 & 3.1 & 7.3 & 4.9 & 2.6 & 13 & 7.1 & 4.3$^b$ & 0.22 \\
35\,\Msun & 4.4$^b$ &    &     &     &     &     &    &     &     &      \\
\hline\hline
\end{tabular}
\smallskip\\
$^a$All numbers here can be multiplied by 1.2 to 1.3 to account for the 
angular momentum carried away by neutrinos \\
$^b$Became a Wolf-Rayet star during helium burning 
\end{table}

\begin{table}
\caption{Pulsar Rotation Rate With Variable Remnant
Mass$^a$\lTab{PSR(m)}}
\smallskip
\begin{tabular}{lccccc}
\hline\hline\noalign{\smallskip}
Mass  & Baryon$^b$  & Gravitational$^c$&   $J(M_{\rm bary}$)&  BE             & Period$^d$ \\
      &($\Msun$)    & ($\Msun$)        & ($\Ep{47}\,\Junit$)  & ($\Ep{53}\,\erg$) & ($\ms$) \\
\hline\noalign{\smallskip}
12\,\Msun & 1.38 & 1.26 & 5.2 & 2.3 & 15  \\
15\,\Msun & 1.47 & 1.33 & 7.5 & 2.5 & 11  \\
20\,\Msun & 1.71 & 1.52 & 14  & 3.4 & 7.0 \\
25\,\Msun & 1.88 & 1.66 & 17  & 4.1 & 6.3 \\
35\,\Msun$^e$ & 2.30 & 1.97 & 41  & 6.0 & 3.0 \\
\hline\hline
\end{tabular}
\smallskip\\
$^a$Assuming a constant radius of $12$ km 
and a moment of inertia $0.35 M R^2$ (Lattimer \& Prakash 2001)\\
$^b$Mass before collapse where specific entropy is $4\,\Sunit$ \\
$^c$Mass corrected for neutrino losses \\
$^d$Not corrected for angular momentum carried away by neutrinos \\
$^e$ Becaame a Wolf-Rayet star during helium burning
\end{table}

\begin{table}
\caption{Periods and Angular Momentum Estimates for Observed Young
Pulsars\lTab{PSR-obs}}
\smallskip
\begin{tabular}{lccc}
\hline\hline\noalign{\smallskip}
                            & current & initial &  J$_o$          \\
\raisebox{1.5ex}[0pt]{pulsar}
                            & (ms)   & (ms)  & (\Junit)   \\
\hline\noalign{\smallskip}
PSR J0537-6910 (N157B, LMC) &  16    & $\sim$10 & 8.8\E{47} \\
PSR B0531+21 (crab) \dotfill&  33    & 21 & 4.2\E{47} \\
PSR B0540-69 (LMC) \dotfill &  50    & 39 & 2.3\E{47} \\
PSR B1509-58  \dotfill      & 150    & 20 & 4.4\E{47} \\
\hline
\end{tabular}
\end{table}

\begin{table}
\caption{Evolution of Surface Abundances ($15\,\Msun$)\lTab{Xsurf}}
\smallskip
\begin{tabular}{l|lccc|lccc}
\hline\hline\noalign{\smallskip}
                            & & H-Burn$^a$ & He-Burn$^b$	 & PreSN &  &
                            H-Burn$^a$   & He-Burn$^b$   & PreSN   \\
\hline\noalign{\smallskip}
no-rot & $^{4}$He  &  0.2762 & 0.2781 & 0.3292 & $^{12}$C  & 2.8\E{-3} &2.3\E{-3} &1.6\E{-3} \\
rot    & $^{4}$He  &  0.2777 & 0.2872 & 0.3401 & $^{12}$C  & 2.2\E{-3} &1.2\E{-3} &9.1\E{-4} \\
rot+B  & $^{4}$He  &  0.2770 & 0.2833 & 0.3365 & $^{12}$C  & 2.2\E{-3} &1.4\E{-3} &1.0\E{-3} \\
\noalign{\smallskip}\hline\noalign{\smallskip}
no-rot & $^{13}$C  & 3.4\E{-5} &1.3\E{-4} & 1.0\E{-4}& $^{15}$N  & 3.2\E{-6} &1.8\E{-6} &1.3\E{-6} \\
rot    & $^{13}$C  & 1.2\E{-4} &1.7\E{-4} & 1.4\E{-4}& $^{15}$N  & 1.6\E{-6} &4.9\E{-7} &4.3\E{-7} \\
rot+B  & $^{13}$C  & 1.3\E{-4} &1.9\E{-4} & 1.5\E{-4}& $^{15}$N  & 1.6\E{-6} &4.9\E{-7} &4.3\E{-7} \\
\noalign{\smallskip}\hline\noalign{\smallskip}
no-rot & $^{14}$N  & 8.2\E{-4} &1.3\E{-3} & 3.3\E{-3}& $^{16}$O  & 7.6\E{-3} &7.6\E{-3} &6.3\E{-3} \\
rot    & $^{14}$N  & 1.6\E{-3} &3.0\E{-3} & 4.4\E{-3}& $^{16}$O  & 7.5\E{-3} &7.0\E{-3} &6.0\E{-3} \\
rot+B  & $^{14}$N  & 1.5\E{-3} &2.6\E{-3} & 4.1\E{-3}& $^{16}$O  & 7.5\E{-3} &7.3\E{-3} &6.1\E{-3} \\
\noalign{\smallskip}\hline\noalign{\smallskip}
no-rot & $^{17}$O  & 3.1\E{-6} &3.6\E{-6} & 5.7\E{-6}& $^{18}$O  & 1.7\E{-5} &1.6\E{-5} &5.7\E{-6} \\
rot    & $^{17}$O  & 5.4\E{-6} &8.2\E{-6} & 8.1\E{-6}& $^{18}$O  & 1.5\E{-5} &1.0\E{-5} &7.6\E{-6} \\
rot+B  & $^{17}$O  & 4.9\E{-6} &7.0\E{-6} & 7.2\E{-6}& $^{18}$O  & 1.5\E{-5} &1.1\E{-5} &8.6\E{-6} \\
\noalign{\smallskip}\hline\noalign{\smallskip}
no-rot & $^{23}$Na & 3.4\E{-5} &3.6\E{-5} & 6.4\E{-5}& $^{19}$F  & 4.1\E{-7} &4.0\E{-7} &3.2\E{-7} \\
rot    & $^{23}$Na & 4.5\E{-5} &6.1\E{-5} & 8.0\E{-5}& $^{19}$F  & 3.7\E{-7} &3.3\E{-7} &2.6\E{-7} \\
rot+B  & $^{23}$Na & 4.2\E{-5} &5.2\E{-5} & 7.5\E{-5}& $^{19}$F  & 3.8\E{-7} &3.5\E{-7} &2.8\E{-7} \\
\noalign{\smallskip}\hline\noalign{\smallskip}
no-rot & $^{11}$B  & 3.8\E{-9} &2.6\E{-10}& 1.8\E{-10}& $^{9}$Be  & 1.7\E{-10}&3.5\E{-13}&2.4\E{-13}\\ 
rot    & $^{11}$B  & 2.4\E{-10}&1.3\E{-11}& 8.4\E{-12}& $^{9}$Be  & 9.8\E{-14}&6.0\E{-16}&4.0\E{-16}\\ 
rot+B  & $^{11}$B  & 1.1\E{-10}&6.5\E{-12}& 4.5\E{-12}& $^{9}$Be  & 2.9\E{-15}&5.7\E{-18}&3.9\E{-18}\\ 
\hline
\end{tabular}
\smallskip\\
NOTE: 
$^a$35\,\% central hydrogen mass fraction; 
$^b$50\% central helium mass fraction; 
\end{table}

\clearpage
\begin{figure}
\newcommand{\panelwidth}{0.48\columnwidth}

\includegraphics[draft=\Draft,angle=90,width=\panelwidth]{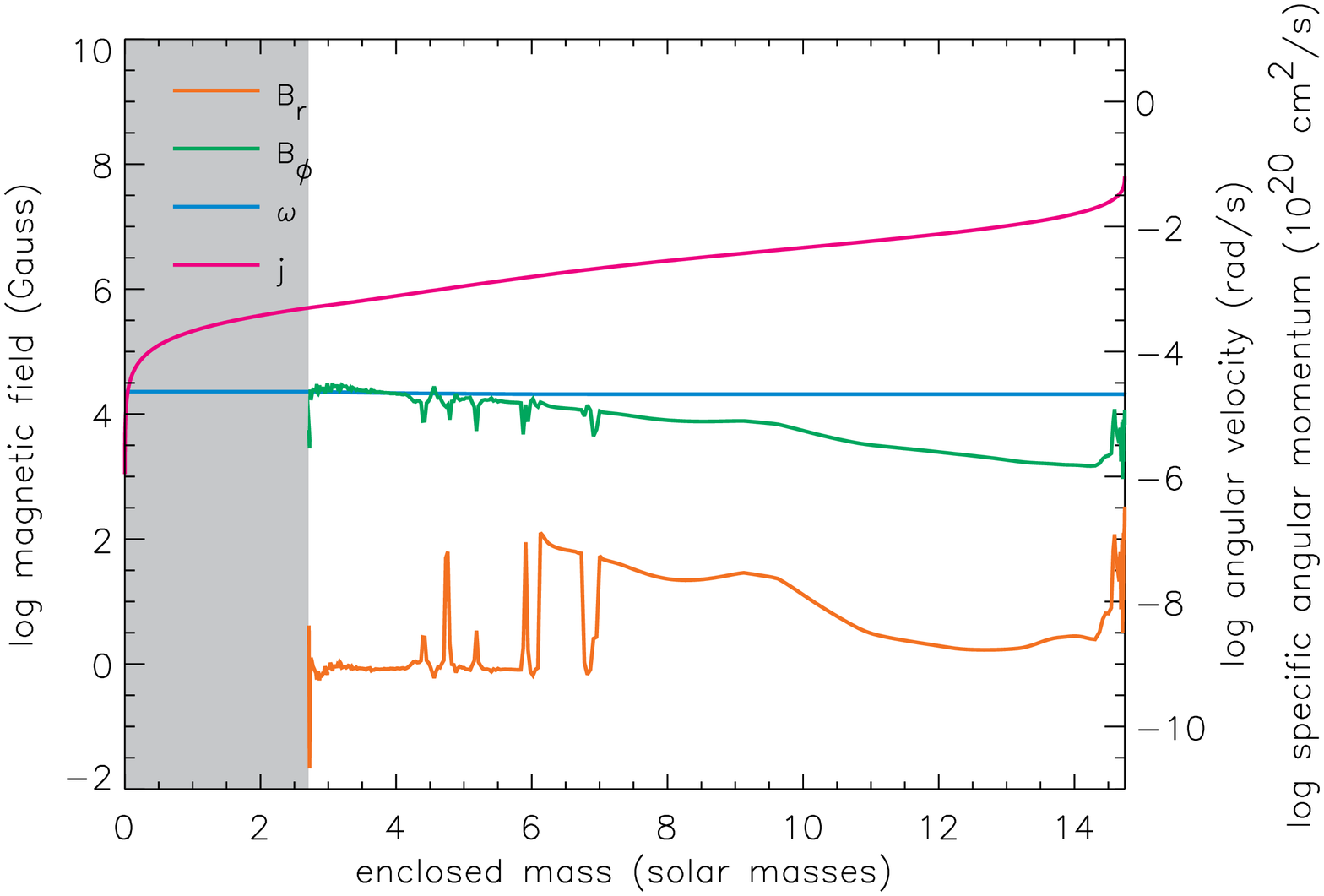}\hfill 
\includegraphics[draft=\Draft,angle=90,width=\panelwidth]{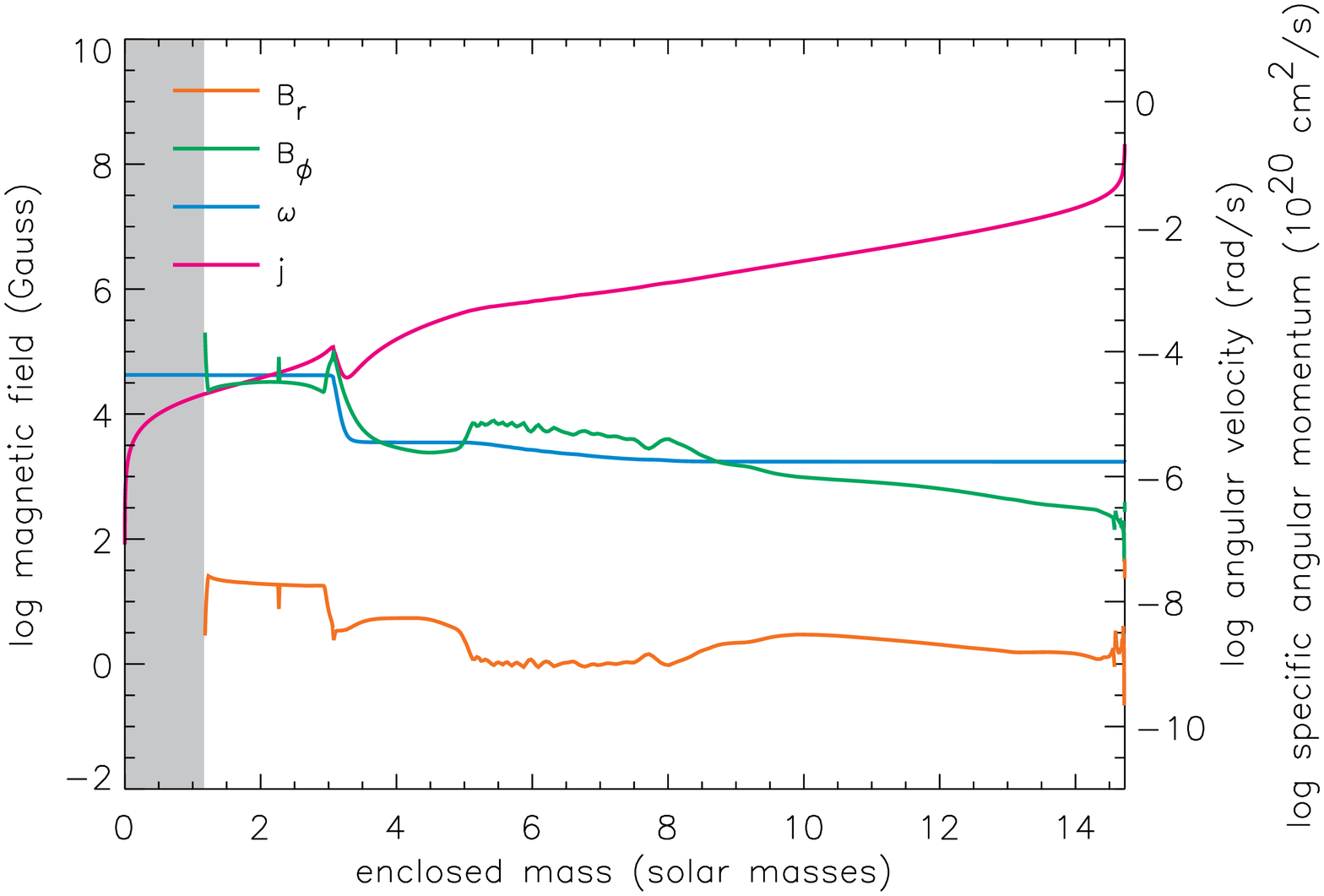}\\[\baselineskip] 
\includegraphics[draft=\Draft,angle=90,width=\panelwidth]{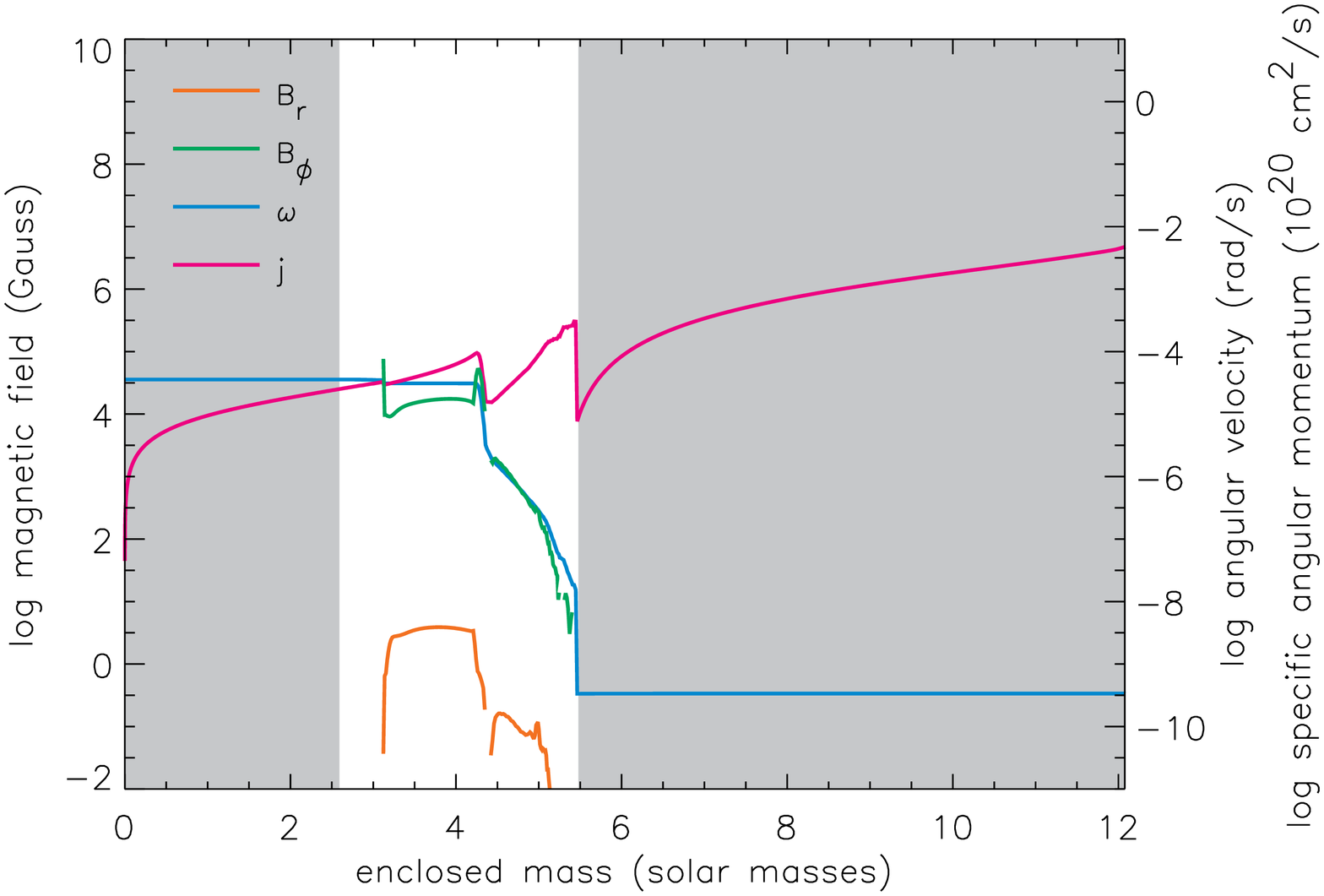}\hfill 
\includegraphics[draft=\Draft,angle=90,width=\panelwidth]{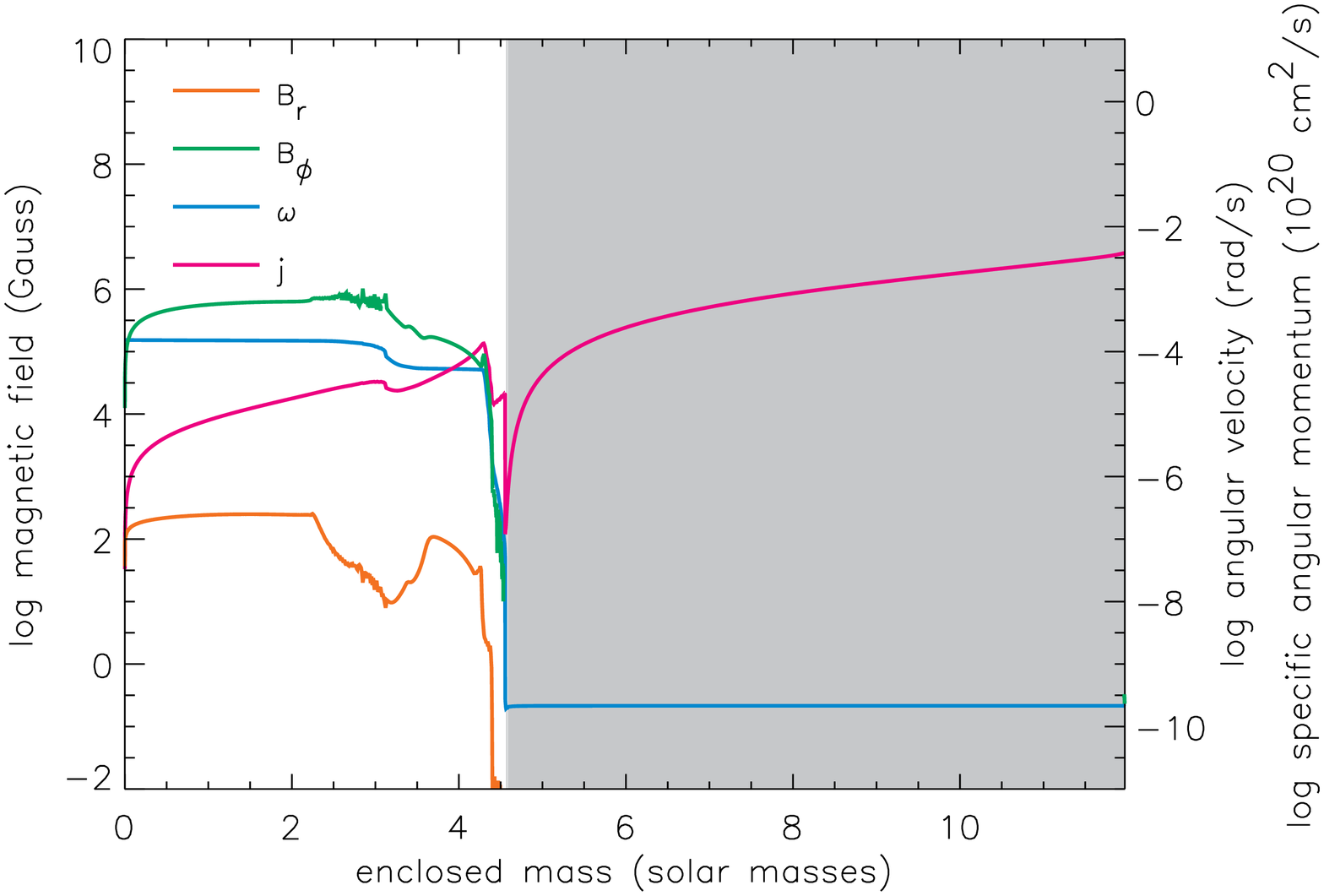}\\ 
\caption{Magnetic field structure and angular momentum distribution
for the standard $15\,\Msun$ model at at hydrogen depletion (upper
left), helium ignition (upper right), helium depletion (lower left),
and carbon ignition (lower right).  See \Tab{15evo} for the definitions
of these times. The shaded regions are those portions of the star that
are convective.  In those regions large diffusion coefficients for
angular momentum lead to nearly rigid rotation in all but the latest
stages of the evolution.  \lFig{bfig}}
\end{figure}

\clearpage
\begin{figure}
\newcommand{\panelwidth}{0.75\columnwidth}
\centering
\includegraphics[draft=\Draft,angle=0,width=\panelwidth]{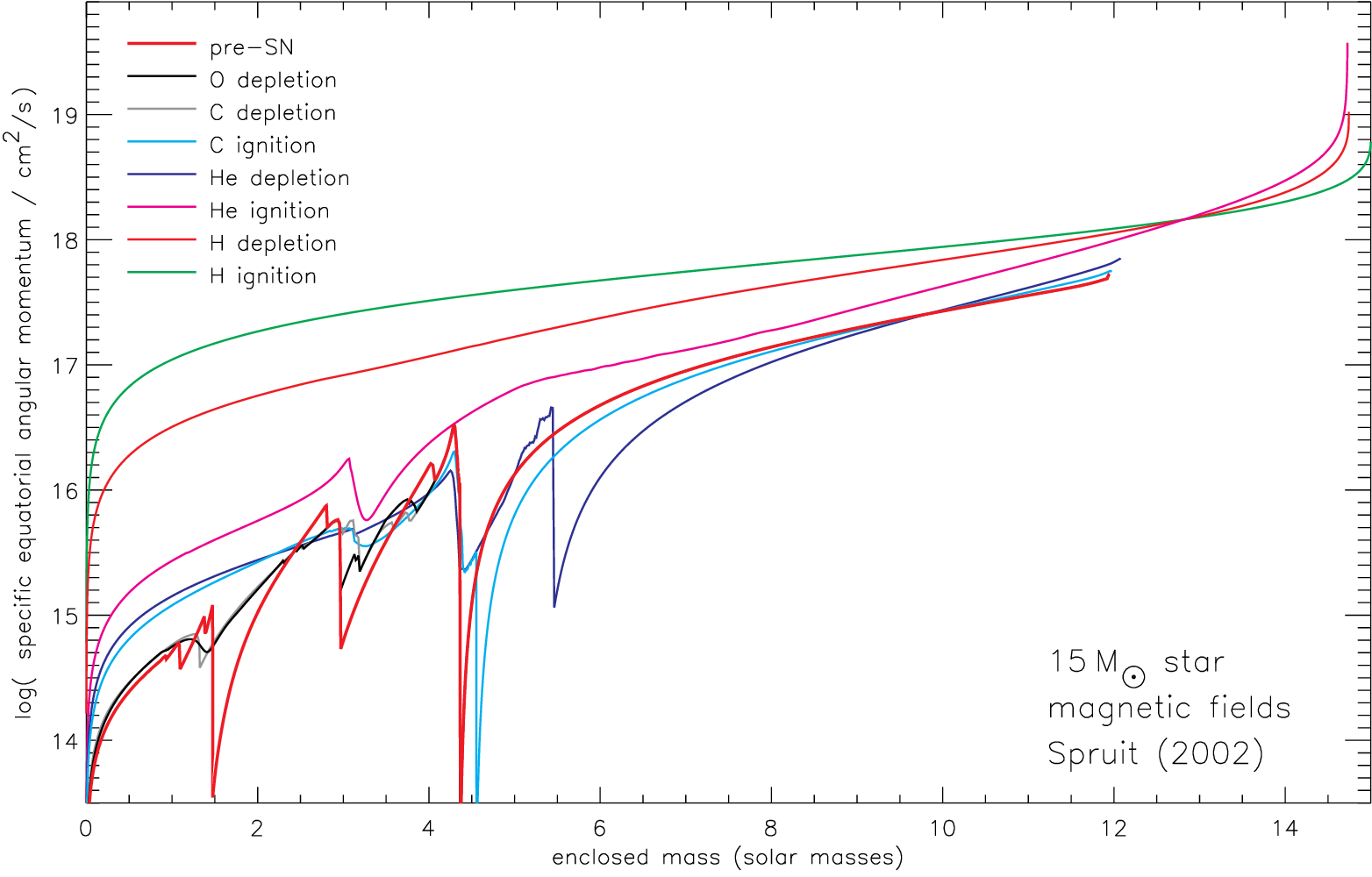}\\ 
\vskip 0.25in
\includegraphics[draft=\Draft,angle=0,width=\panelwidth]{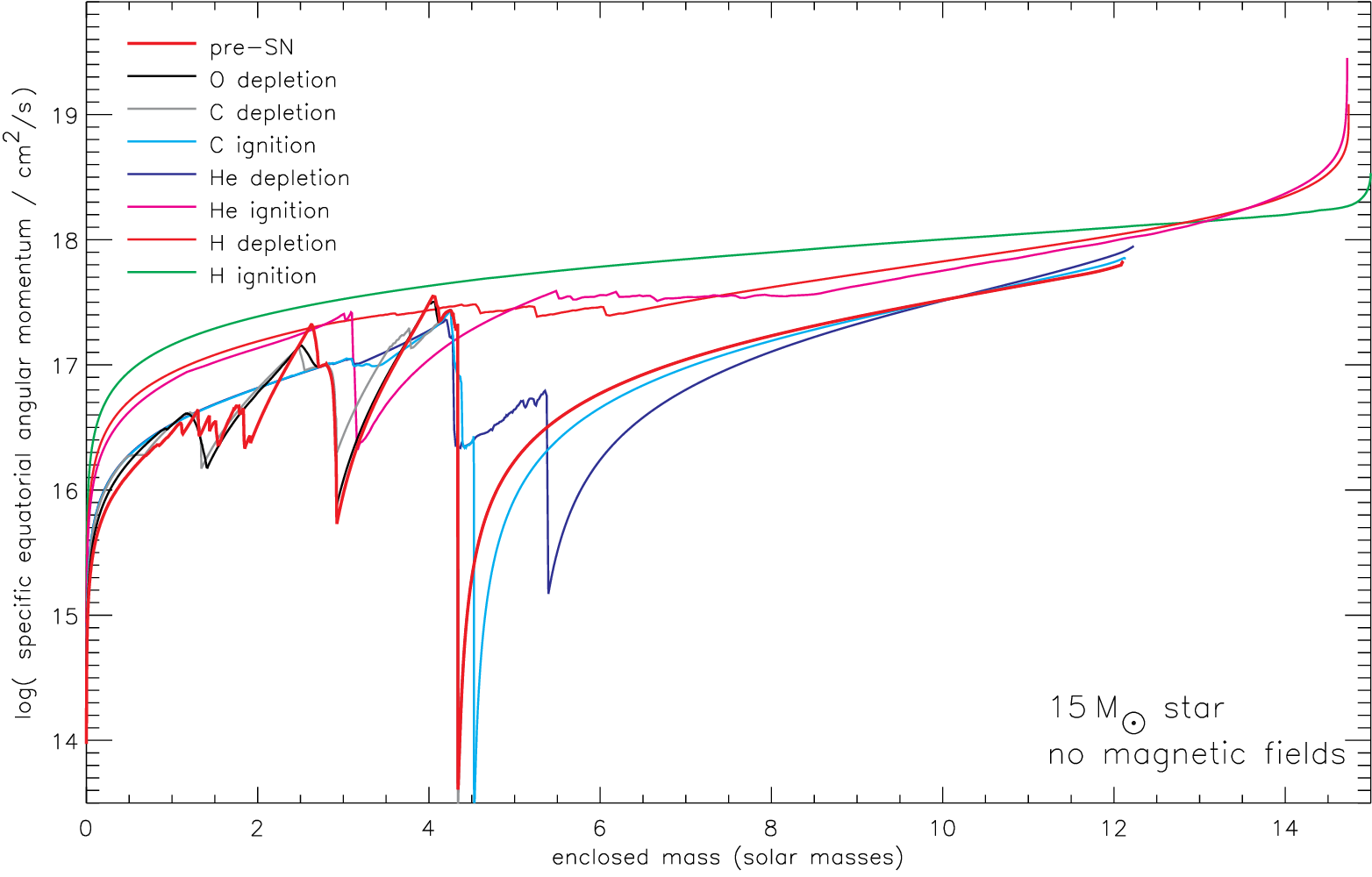}\\[\baselineskip] 
\caption{Specific angular momentum distribution for the standard
$15\,\Msun$ model with (upper panel) and without (lower panel)
magnetic fields at different evolution stages.  See \Tab{15evo} for
the definitions of these times. \lFig{jfig15}}
\end{figure}

\clearpage
\begin{figure}
\newcommand{\panelwidth}{0.75\columnwidth}
\centering
\includegraphics[draft=\Draft,angle=0,width=\panelwidth]{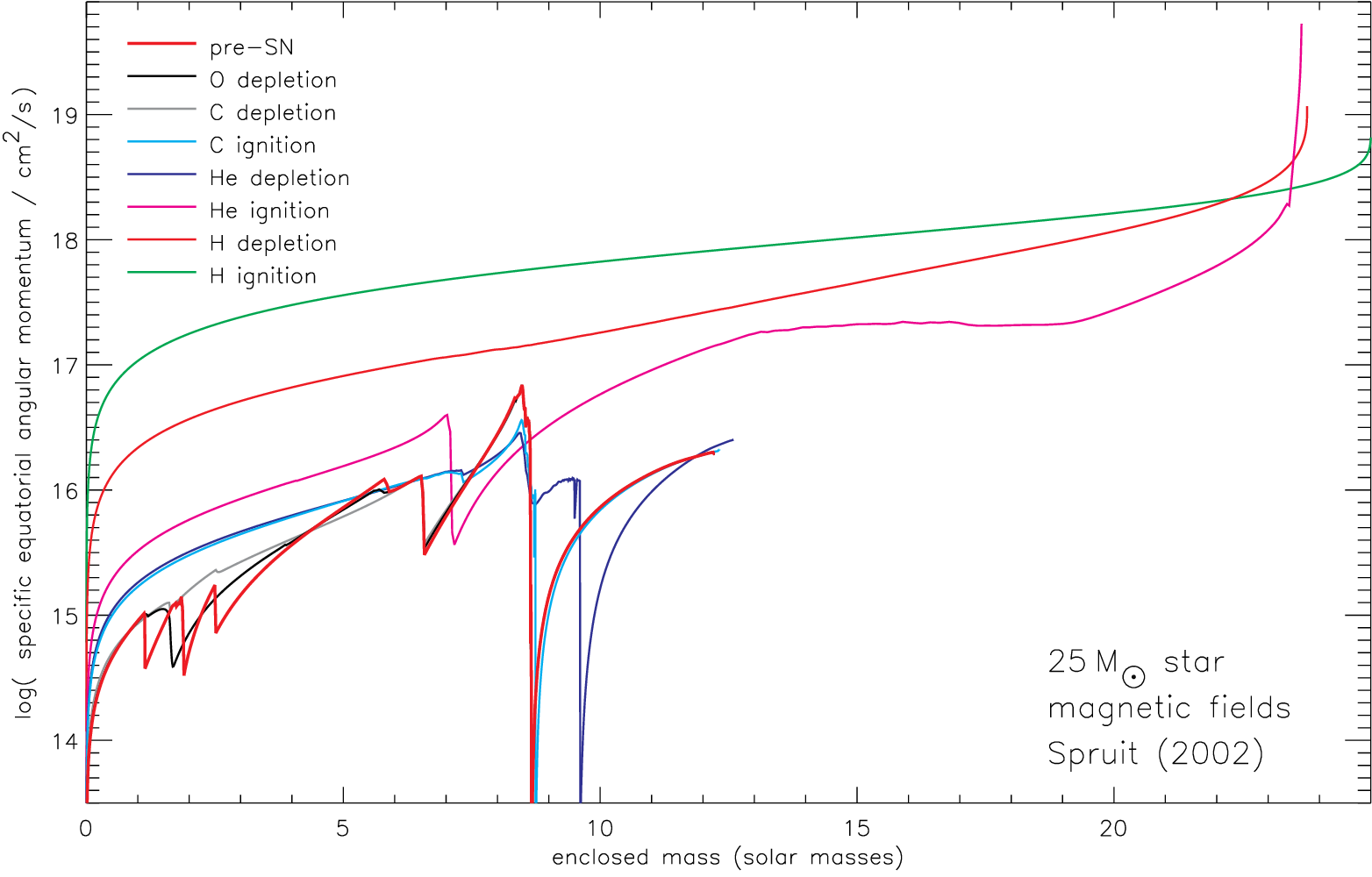}\\ 
\vskip 0.25in
\includegraphics[draft=\Draft,angle=0,width=\panelwidth]{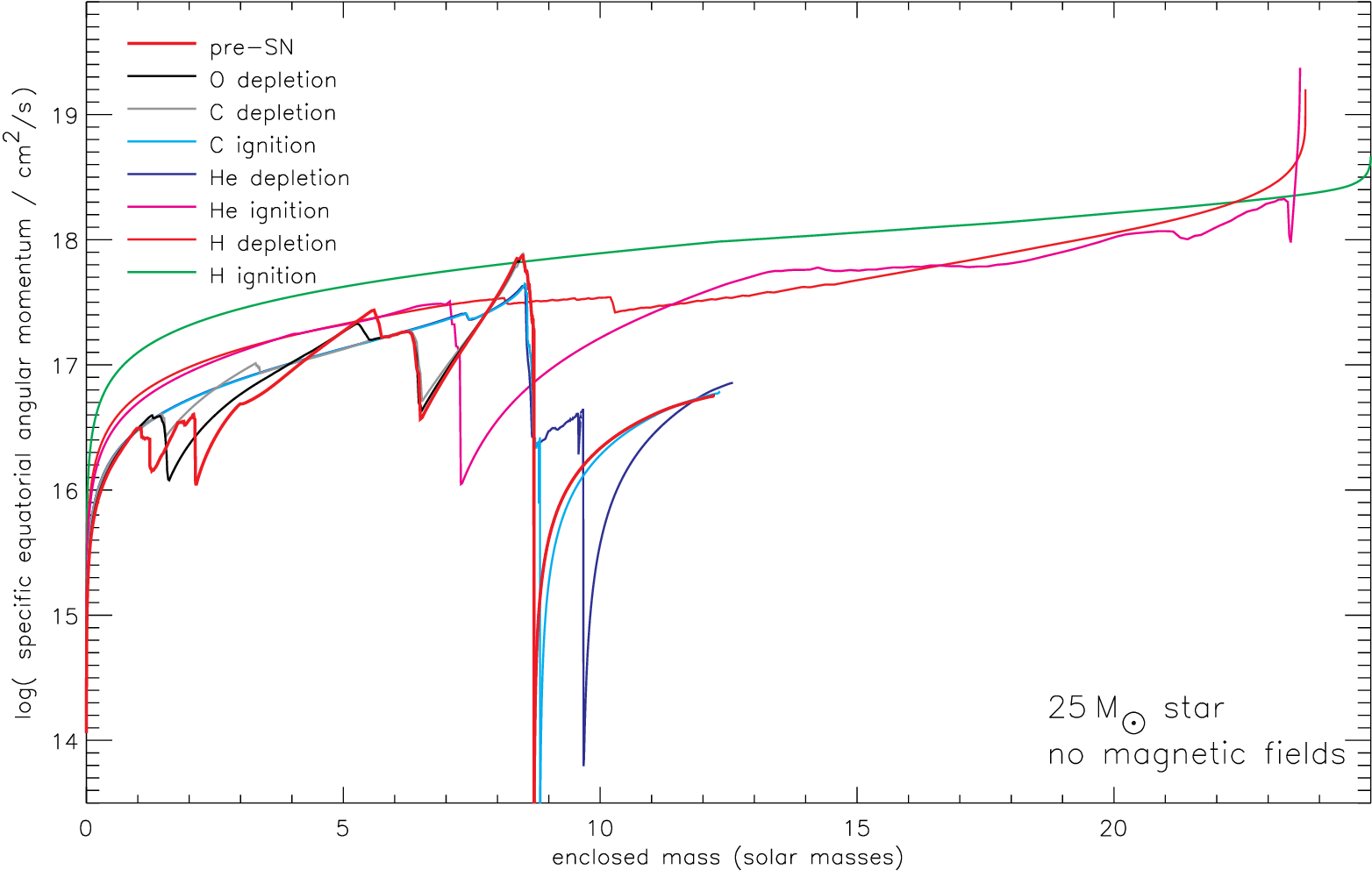}\\[\baselineskip] 
\caption{Same as \Fig{jfig15} but for $25\,\Msun$ stars. \lFig{jfig25}}
\end{figure}

\end{document}